\documentclass[12pt]{article}
\usepackage[margin=1in]{geometry}
\usepackage{setspace}
\usepackage{titling}
\usepackage{color}
\usepackage{colortbl}
\usepackage{fancyvrb}
\usepackage{graphicx}

\usepackage[bf,margin=20pt,format=hang,font={small,stretch=1.1}]{caption}
\usepackage{tikz}
\usetikzlibrary{math}
\usepackage{amsmath,amssymb}
\makeatletter\let\over\@@over\makeatother    
\usepackage{hyperref}
\usepackage{subcaption}
\usepackage{diagbox}

\newcommand\TL{\hfil$\displaystyle{##}$}
\newcommand\TR{$\displaystyle{{}##}$\hfil}

\newcommand\TT{\hbox{##}}
\def\seqalign#1#2{\vcenter{\openup1\jot
  \halign{\strut #1\cr #2 \cr}}}
\def\lbldef#1#2{\expandafter\gdef\csname #1\endcsname {#2}}
\newcommand{\eqn}[3][]{\lbldef{#2}{(\ref{#2})}%
\begin{equation} \eqalign{#3} \label{#2} \end{equation}}
\def\eqalign#1{\vcenter{\openup1\jot
    \halign{\strut\span\TL & \span\TR\cr #1 \cr
   }}}
\def\eno#1{(\ref{#1})}
\def\href#1#2{#2}
\def\mop#1{\mathop{\rm #1}\nolimits}
\def\sgn{\mop{sgn}}
\def\mod#1{\;(\mop{mod}#1)}
\def\tr{\mop{tr}}
\def\stab{\mop{stab}}
\def\orb{\mop{orb}}

\renewenvironment{abstract}
 {\normalsize
  \begin{center}
   \bfseries \abstractname\vspace{-.5em}\vspace{0pt}
  \end{center}
  \list{}{
   \setlength{\leftmargin}{0in}%
   \setlength{\rightmargin}{\leftmargin}%
  }%
  \item\relax}
 {\endlist}

\title{Higher melonic theories}

\author{Steven S.~Gubser, Christian Jepsen, Ziming Ji, and Brian Trundy}
\date{}

\begin{document}
\VerbatimFootnotes  
\begin{titlingpage}

\setlength{\droptitle}{-70pt}
\maketitle
\begin{abstract}
We classify a large set of melonic theories with arbitrary $q$-fold interactions, demonstrating that the interaction vertices exhibit a range of symmetries, always of the form $\mathbb{Z}_2^n$ for some $n$, which may be $0$.  The number of different theories proliferates quickly as $q$ increases above $8$ and is related to the problem of counting one-factorizations of complete graphs.  The symmetries of the interaction vertex lead to an effective interaction strength that enters into the Schwinger-Dyson equation for the two-point function as well as the kernel used for constructing higher-point functions.
\end{abstract}
\vfill
June 2018
\end{titlingpage}

\tableofcontents

\clearpage

\section{Introduction}
\label{INTRODUCTION}

\begin{figure}[b]
  \centerline{\includegraphics[width=3in]{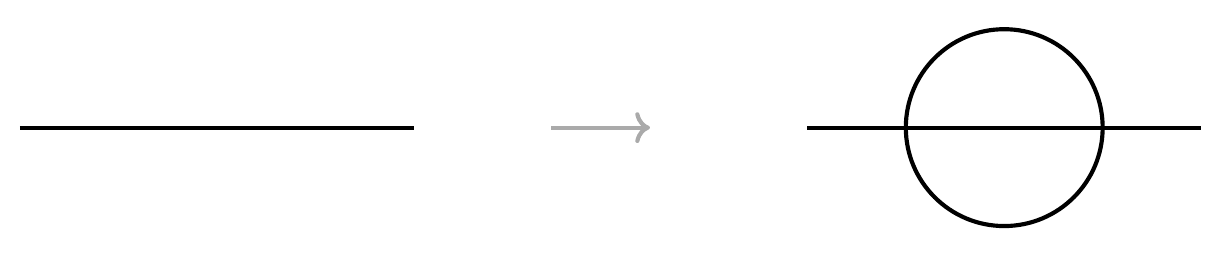}}
  \caption{A melonic insertion.}\label{MelonicInsertion2}
 \end{figure}

Melonic theories \cite{Bonzom:2011zz,Gurau:2011xp,Carrozza:2015adg,Witten:2016iux,Klebanov:2016xxf} are an interesting class of quantum field theories whose essential property is that in an appropriate large $N$ limit, the dominant Feynman diagrams can be generated by iterating on the replacement of a propagator by a melonic insertion, as shown in figure~\ref{MelonicInsertion2} for a melonic version of scalar $\phi^4$ theory.  Melonic theories are interesting for two related reasons: 1) The melonic large $N$ limit is relatively tractable because its Green's functions can be determined through functional techniques including Schwinger-Dyson equations; and 2) The simplest Green's functions are the same as for the Sachdev-Ye-Kitaev (SYK) model \cite{Sachdev:1992fk,Kitaev:2015zz}, widely studied because of its proposed relationship to $AdS_2$.

It is well recognized that melonic theories theories exist not just with quartic interactions, but also higher order interactions \cite{Narayan:2017qtw,Choudhury:2017tax,Ferrari:2017jgw,Tarnopolsky:2018env}.  However, in the literature to date there is not much systematic understanding of what sort of higher order interactions are possible.  The aim of this article is to take some steps toward such an understanding in the context of higher order versions of the Klebanov-Tarnopolsky model \cite{Klebanov:2016xxf} where the interaction is of order $q$---meaning that $q$ propagators meet at each interaction vertex.  Inquiries in this direction were initiated in \cite{prethesis}.  Starting at $q=8$, the number of different interaction vertices proliferates quickly.  Some of them are symmetrical under a subgroup of permutations of the propagators leading into them; other have no such symmetry.  Interaction vertices do not mix with one another in the leading melonic limit: The diagrams that would permit this are subleading.  We are therefore content to restrict to theories with only one type of interaction vertex---and each different interaction vertex gives a different theory.  An interaction vertex of order $q$ can be constructed starting from a coloring of the complete graph with $q$ vertices such that each of $q-1$ colors is incident once on each vertex.  The questions we wish to settle regarding these interaction vertices are:
 \begin{enumerate}
  \item What are the possible symmetry groups of these interaction vertices?
  \item How does the number of distinct interaction vertices grow with $q$?
 \end{enumerate}
We claim in sections~\ref{SYMMETRY} and~\ref{CONSTRUCTION} to completely settle question \#1: The possible symmetry groups are $\mathbb{Z}_2^n$, where $0 \leq n \leq v$ if $q = 2^v$ and $0 \leq n < v$ if $q = u 2^v$ where $u$ is odd and larger than~$1$.  (It is easy to show that $q$ must be even.)  Our demonstration is constructive, in that we produce vertices with each possible symmetry.  Question \#2 turns out to be difficult, and it is essentially the problem of counting so-called one-factorizations of complete graphs, where results are generally available up to $q=14$ \cite{Dickson:1906,Gelling:1973,Gelling:1974,Dinitz:1987,Seah,Dinitz:1994,Kaski:2009}: results spanning over a century!  In section~\ref{ONEFACTOR}, we summarize how these results can be combined and modestly extended to give complete results on the number of distinct interaction vertices up to $q=14$.  Our symmetry analysis suggests a new twist on the counting problem: In addition to counting all one-factorizations, one can count one-factorizations with a given symmetry group.  An explicit example of this symmetry-constrained counting is presented in appendix~\ref{AlgorithmAppendix}, and some additional conceptual points are discussed in appendix~\ref{IsomorphismOrdered}.

The symmetry group of the interaction vertex leads to an effective coupling that enters into the Schwinger-Dyson equation for the two-point function and the ladder operator used for computing the four-point function, which we exhibit explicitly in all cases in sections~\ref{TWOPOINT} and~\ref{FOURPOINT}.\footnote{All indications are that every interaction vertex we construct leads to a theory with a melonic limit; however, we do not have a fully rigorous proof of this claim.}  We work out all our results for theories not only over the reals $\mathbb{R}$, but also over the $p$-adic numbers $\mathbb{Q}_p$, where as already shown in \cite{Gubser:2017qed}, there is quite a variety of melonic theories, depending on what sign function one chooses over $\mathbb{Q}_p$.  For approximately half of these melonic theories over $\mathbb{Q}_p$, the Schwinger-Dyson equation can be solved exactly, not just in the infrared, but at all scales, in terms of the solution to a $q$-th order polynomial equation.  Remarkably, as shown in section~\ref{ADELIC}, there is an adelic product formula relating the eigenvalues of the ladder operator integral equation across real and $p$-adic theories.

\section{Structure of higher melonic theories}
\label{STRUCTURE}

The action of the simplest Klebanov-Tarnopolsky model \cite{Klebanov:2016xxf} is
 \eqn{SKT}{
  S = \int_{\mathbb{R}} dt \, \left[ 
    {i \over 2} \psi^{a_0a_1a_2} \partial_t \psi^{a_0a_1a_2} + 
    {g \over 4} \psi^{a_0a_1a_2} \psi^{a_0b_1b_2} \psi^{c_0a_1b_2} \psi^{c_0b_1a_2} \right] \,,
 }
where each index $a_0$, $a_1$, $a_2$, which we can think of as a color index, runs independently from $1$ to $N$, and each $\psi^{a_0a_1a_2}$ is a Majorana fermion.    
There is a different $O(N)$ for each index position, so that the whole of $\psi^{a_0a_1a_2}$ transforms in the tensor product of the fundamental representations of three copies of $O(N)$. 
We are interested in the color structure for higher rank models with higher degree interactions. We also want to generalize to models defined over the $p$-adic numbers $\mathbb{Q}_p$ and to models with $O(N)$ or $Sp(N)$ indices, as in \cite{Gubser:2017qed}.  Schematically, the lagrangians we will consider take the form
 \eqn{SKTextended}{
  S &= \sigma_\psi 
   \int_K d\omega \, {1 \over 2} \psi^A(-\omega) \psi^B(\omega) |\omega|^s (\sgn\omega) 
    \Omega_{AB}  \cr
    &\qquad\qquad{} + (\sigma_\psi)^{\frac{q}{4}} 
   \int_K dt \, {g \over |G|} \left( \prod_{i=0}^{q-1} \psi^{A^{(i)}}(t) \right)
    \Omega_{A^{(0)}A^{(1)}\dots A^{(q-1)}} \,,
 }
where $K$ is either $\mathbb{R}$ or $\mathbb{Q}_p$ (or possibly some larger vector space, for instance $\mathbb{C}$ or a field extension of some $\mathbb{Q}_p$) and $G$ is the automorphism group of the interaction vertex, to be discussed further in section~\ref{SYMMETRY}.  The spectral parameter $s$ would usually be chosen to be $1$ for fermionic theories over $\mathbb{R}$ or $2$ for bosonic theories over $\mathbb{R}$, but for theories over $\mathbb{Q}_p$ it is more natural to let it vary continuously over positive real values.  We set $\sigma_\psi = -1$ for fermionic theories, and $\sigma_\psi = +1$ for bosonic theories.  By $\sgn\omega$ we mean a sign character, which is to say a multiplicative homomorphism of non-zero elements of $K$ to $\{1,-1\}$.  Capital indices are really groups of $q-1$ lowercase indices, each $N$-valued.  For example, to recover \eno{SKT} as a special case, we would set $q=4$, so that $A = a_0a_1a_2$; we would set $\Omega_{AB} = \delta_{a_0b_0} \delta_{a_1b_1} \delta_{a_2b_2}$; we would set
 \eqn{OmegaSimple}{
  \Omega_{ABCD} = \delta_{a_0b_0} \delta_{a_1c_1} \delta_{a_2d_2} 
    \delta_{b_1d_1} \delta_{b_2c_2} \delta_{c_0d_0} \,;
 }
and of course we would set $K=\mathbb{R}$ and $s=1$.  It is well recognized (see e.g.~\cite{Gurau:2011xp,Witten:2016iux,Klebanov:2016xxf}) that the structure $\Omega_{ABCD}$ in \eno{OmegaSimple} corresponds to a coloring of the edges of the tetrahedron so that only three colors are used, and opposite edges have the same color.\footnote{It is useful to clarify here one point of terminology: We use the terms ``vertex'' and ``edge'' to describe the inner structure of an interaction vertex like $\psi^{a_0a_1a_2} \psi^{a_0b_1b_2} \psi^{c_0a_1b_2} \psi^{c_0b_1a_2}$.  From this point of view, an interaction vertex is a graph unto itself, with $q$ vertices when the interaction term has $q$ powers of $\psi$.  A full Feynman diagram consists of propagators connecting interaction vertices, and as is familiar from earlier work including \cite{Gurau:2011xp}, the inner structure of a propagator is $q-1$ threads which flow into the edges inside an interaction vertex.} 

When considering theories over $\mathbb{Q}_p$, as explained in \cite{Gubser:2017qed}, we must allow $\Omega_{AB}$ to be symmetric ($\sigma_\Omega = 1$) or anti-symmetric ($\sigma_\Omega = -1$); and we must choose $\sgn\omega$ to be one of the several multiplicative sign characters over $\mathbb{Q}_p$, which are in one-to-one correspondence with the quadratic extensions of $\mathbb{Q}_p$.  To get a real, non-vanishing kinetic term, $\Omega_{AB}$ must be Hermitian, and we must have
 \eqn{sgnConstraint}{
  \sigma_\psi \sigma_\Omega = \sgn(-1) \,.
 }
(Surprisingly, non-trivial sign characters over $\mathbb{Q}_p$ can have either $\sgn(-1) = -1$ or $+1$.)

For $q=4$, still following \cite{Gubser:2017qed}, the obvious adaptation of the Klebanov-Tarnopolsky model to a theory over $\mathbb{Q}_p$ is to set $\Omega_{AB} = \Omega_{a_0b_0} \Omega_{a_1b_1} \Omega_{a_2b_2}$ where
 \eqn{OmegaLowercase}{
  \Omega_{ab} = \left\{ \seqalign{\span\TR &\qquad\hbox{for}\qquad \span\TR}{
   {\bf 1}_{N \times N} & \sigma_\Omega = 1  \cr
   \sigma_2 \otimes {\bf 1}_{{N \over 2} \times {N \over 2}} & \sigma_\Omega = -1 \,,} \right.
 }
and to set
 \eqn{OmegaInt}{
  \Omega_{ABCD} = 
    \Omega_{a_0b_0} \Omega_{a_1c_1} \Omega_{a_2d_2} \Omega_{b_1d_1} \Omega_{b_2c_2} \Omega_{c_0d_0}
    \,.
 }
(Note that if $\sigma_\Omega = -1$, then because of \eno{OmegaLowercase}, $N$ must be even.)  If indeed $\Omega_{ab}$ is antisymmetric, then one needs a direction on all the edges of the tetrahedral graph in order to decide the order of the indices in each factor on the right hand side of \eno{OmegaInt}.  However, flipping the direction on any one edge flips the sign of $\Omega_{ABCD}$, and so can be compensated for by changing the sign of $g$.

Generalizing the kinetic term to $q>4$ is easy: We need only set
 \eqn{OmegaABgen}{
  \Omega_{AB} = \prod_{i=0}^{q-2} \Omega_{a_ib_i} \,.
 }
Generalizing the interaction tensor turns out to be more subtle, and laying the groundwork for finding suitable generalizations is the focus of the rest of this section.  

Up to the minor issue of directedness, constructing a rank $q$ interaction tensor $\Omega_{A^{(0)}A^{(1)}\dots A^{(q-1)}}$ as a product of $q(q-1)/2$ factors $\Omega_{a^{(i)}_r a^{(j)}_r}$ corresponds to a coloring problem on the complete graph of $q$ points (and therefore $q(q-1)/2$ edges), where we use $q-1$ colors (each one labeled by a value of $r$) and require that each of the $q-1$ links incident on a given vertex (each one labeled by a value of $i$) must be a different color.  A special case of Baranyai's theorem guarantees that this can always be done provided $q$ is even.  It is impossible when $q$ is odd.  One can map the problem onto the scheduling of a round-robin tournament, where each link is one game, each vertex is a contestant, and each color is a round, during which each contestant plays exactly one game.  This phrasing makes it obvious that no coloring is possible for $q$ odd, because in a given round one must pair up all $q$ contestants in two-person games.  For $q$ even, there is a canonical solution, which is
 \eqn{rnm}{\seqalign{\span\TL &\qquad\span\TT}{
  r_{ij} \equiv i+j \mod{q-1} & if $0 \leq i < q-1$, $0 \leq j < q-1$, and $i \neq j$ \cr
  r_{i,q-1} \equiv 2i \mod{q-1} & for $0 \leq i < q-1$\,.
 }}
Here, $r_{ij}$ is the color of the edge from vertex $i$ to vertex $j$.  Vertex labels $i$ and $j$ take values from $0$ to $q-1$, while color labels $r$ run from $0$ to $q-2$.  The corresponding interaction tensor is
 \eqn{InteractionTensor}{
  \Omega_{A^{(0)}A^{(1)}\dots A^{(q-1)}} = \prod_{0 \leq i < j \leq q-1}
    \Omega_{a^{(i)}_{r_{ij}} a^{(j)}_{r_{ij}}} \,.
 }
In \eno{InteractionTensor}, we took care of the directedness issue by requiring $i<j$, which is the same as alphabetizing the lowercase indices in \eno{OmegaInt}.

 \begin{figure}
    \centering

\begin{subfigure}[b]{0.3\textwidth}
        \includegraphics[width=\textwidth]{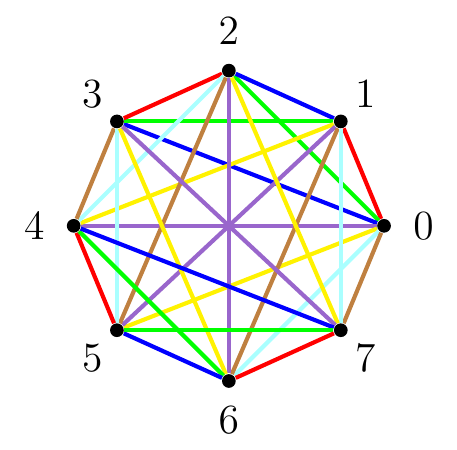}
        \caption{\hspace{12mm}$G=\mathbb{Z}_2^{\,3}$\\30 one-factorizations  \\Order of $\mathrm{Aut}(\overline{\mathcal{F}})$ is 1344}
        \label{fig:oct30}
    \end{subfigure}
    \begin{subfigure}[b]{0.3\textwidth}
        \includegraphics[width=\textwidth]{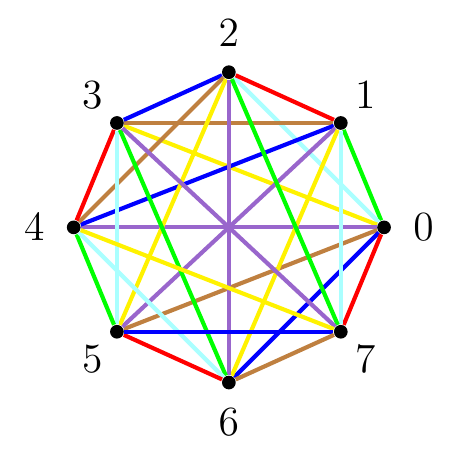}
        \caption{\hspace{12mm}$G=\mathbb{Z}_2^{\,2}$\\630 one-factorizations  \\Order of $\mathrm{Aut}(\overline{\mathcal{F}})$ is 64}
        \label{fig:oct630}
    \end{subfigure}
    \begin{subfigure}[b]{0.3\textwidth}
\includegraphics[width=\textwidth]{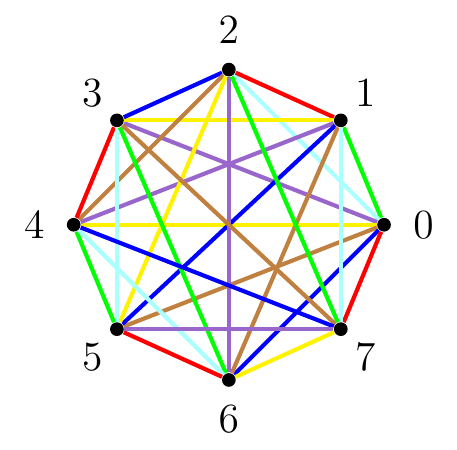}
        \caption{\hspace{12mm}$G=\mathbb{Z}_2^{\,2}$\\420 one-factorizations  \\Order of $\mathrm{Aut}(\overline{\mathcal{F}})$ is 96}
        \label{fig:oct420}
\end{subfigure}

    \begin{subfigure}[b]{0.3\textwidth}
        \includegraphics[width=\textwidth]{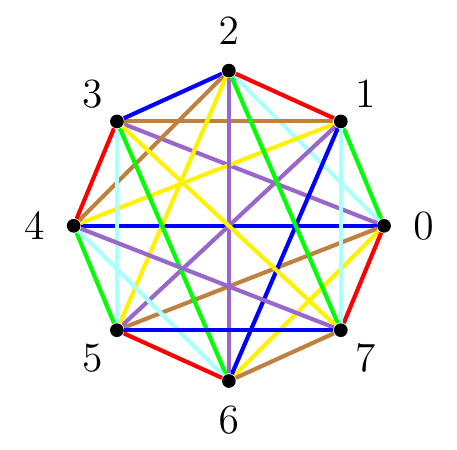}
        \caption{\hspace{12mm}$G=\mathbb{Z}_2$\\2520 one-factorizations  \\Order of $\mathrm{Aut}(\overline{\mathcal{F}})$ is 16}
        \label{fig:oct2520}
    \end{subfigure}
    \begin{subfigure}[b]{0.3\textwidth}
        \includegraphics[width=\textwidth]{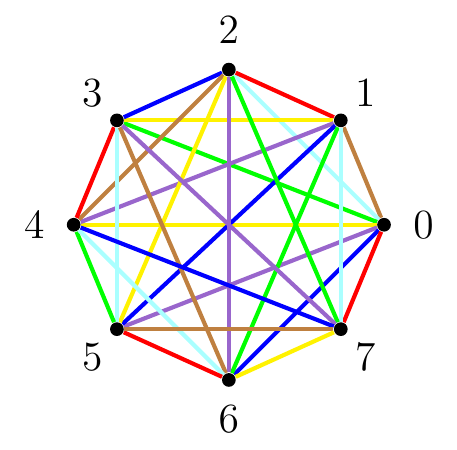}
        \caption{\hspace{12mm}$G=1$\\1680 one-factorizations  \\Order of $\mathrm{Aut}(\overline{\mathcal{F}})$ is 24}
        \label{fig:oct1680}
    \end{subfigure}
    \begin{subfigure}[b]{0.3\textwidth}
\includegraphics[width=\textwidth]{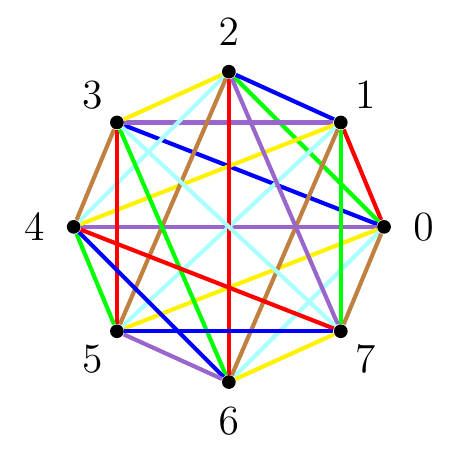}
        \caption{\hspace{12mm}$G=1$\\960 one-factorizations  \\Order of $\mathrm{Aut}(\overline{\mathcal{F}})$ is 42}
        \label{fig:oct960}
\end{subfigure}

\caption{The six inequivalent melonic interactions and their symmetry factors for $q=8$. Each interaction can be identified with an isomorphism class of one-factorizations, and we also list the number of one-factorizations in each class as well as $|\mathrm{Aut}(\overline{\mathcal{F}})|$, the number of permutations in $S_8$ that preserve the one-factorizations in a given isomorphism class (see section \ref{ONEFACTOR} for an explanation of this notation). The canonical coloring is the bottom right.}
    \label{Octagons}
\end{figure}

For $q=2$, $4$, and $6$, the canonical solution \eno{rnm} is the only solution.  Starting at $q=8$, there are multiple solutions: That is, $r_{ij}$ can be chosen differently from \eno{rnm} but still consistent with the requirement that we use only $q-1$ colors and have one edge of each color coming together at each vertex.  For any such $r_{ij}$, we can still use \eno{InteractionTensor} to construct the interaction tensor.  We exhibit the six different solutions for $q=8$ in figure~\ref{Octagons}.  By ``different,'' we mean that there is no way to relabel the colors and/or the vertices to map any of the six $r_{ij}$ into one another.  A striking point is that the solutions have different symmetry groups, composed of up to three factors of $\mathbb{Z}_2$.\footnote{By $\mathbb{Z}_2$ we mean the integers modulo $2$, or equivalently the multiplicative group $\{1,-1\}$---not the $2$-adic integers.}

For $q=10$, there are (we claim) $396$ different interaction vertices, and none of them have any symmetry.  To justify this claim, and to proceed to larger $q$, we need to give a more conceptually organized presentation.  We do so in the next three sections, starting with constraints on the symmetry group in section~\ref{SYMMETRY}, continuing with an explicit construction in section~\ref{CONSTRUCTION}, and concluding with a summary of the problem of counting distinct interaction vertices in section~\ref{ONEFACTOR}.

\section{Symmetry groups of interaction vertices}
\label{SYMMETRY}

\begin{figure}
\centering{
\includegraphics[height=20ex]{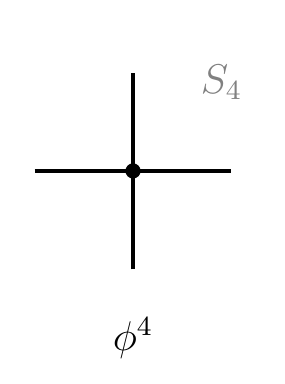} \hspace{6em}
\includegraphics[height=20ex]{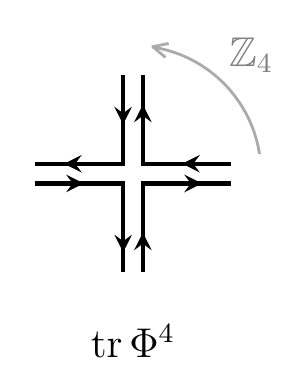}\hspace{4em}
\includegraphics[height=20ex]{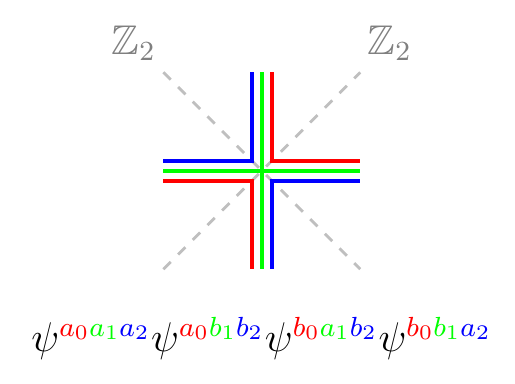}
}
\caption{Left: The symmetry group of $\phi^4$ theory is the symmetric group $S_4$.  Middle: The symmetry group of $\tr \Phi^4$ theory is the cyclic group $\mathbb{Z}_4$.  Right: The symmetry group of the quartic tensor interaction is the Klein group $\mathbb{Z}_2^2$.}\label{VertexGroups}
\end{figure}

In ordinary scalar field theory where the scalar $\phi$ is real-valued, the symmetry group of a $q$-fold interaction vertex $\phi^q$ is the permutation group $S_q$, with order $q!$, because all propagators leading into the interaction vertex are equivalent and can be permuted arbitrarily without changing the structure of the interaction.  In a matrix field theory based on a Hermitian $N \times N$ matrix $\Phi$, the symmetry group of a $\tr \Phi^q$ interaction vertex is the group $\mathbb{Z}_q$ of cyclic permutations of the propagators.  In the $q=4$ Klebanov-Tarnopolsky model, the symmetry group is $\mathbb{Z}_2 \times \mathbb{Z}_2$ (not $\mathbb{Z}_4$), generated by the permutations $(12)(34)$, and $(13)(24)$ and sometimes referred to as the Klein group.  As should be clear from figure~\ref{VertexGroups}, a permutation in the Klein group reorders propagators leading into the interaction vertex in such a way that we get back to exactly the same diagram that we started with.  This sort of permutation is what we will call a coloring automorphism.  If we look at the inner structure of the interaction vertex, we see that a coloring automorphism permutes the vertices (each one corresponding to an incoming propagator) in such a way as to preserve the colors of each edge.

To be precise: When we say that $r_{ij}$ is a coloring of the complete graph $K_q$ with $q$ vertices, what we mean is that $ij$ labels an edge (so $i \neq j$), and each $r_{ij}$ is chosen from the set of ``colors'' $\{0,1,\ldots,q-2\}$, with the constraints that $r_{ij} = r_{ji}$ and that for fixed $i$, $r_{ij}$ is a bijection from the $q-1$ vertices that remain after $i$ is omitted to the set of colors.  In other words, each edge leading into a given vertex is a different color.  A coloring automorphism is defined as a map $i \to \pi(i)$ such that $r_{\pi(i)\pi(j)} = r_{ij}$ for all $i$ and $j$.  Let the group of coloring automorphisms be $G$.  The main purpose of the rest of this section is to limit the possibilities for $G$.  Then in section~\ref{CONSTRUCTION} we will show that all groups $G$ not ruled out by the arguments of this section actually can be realized.

Our first claim is that any coloring automorphism is an involution.  Denote the coloring automorphism by $\pi$.  Assume that $\pi$ is not the identity, since otherwise the claim is trivial.  For some vertex $i$, we have $j \equiv \pi(i) \neq i$.  Then $r_{ij} = r_{\pi(i)\pi(j)} = r_{j\pi(j)}$, where in the first equality we remembered that $\pi$ is a coloring automorphism.  From $r_{ji} = r_{j\pi(j)}$ we can conclude that $i = \pi(j)$ because, as noted previously, the coloring $r$ must be a bijection, for fixed $j$, from vertices $i \neq j$ to colors.

Next we remember an elementary result of group theory: Any finite group $G$ consisting only of involutions is isomorphic to $\mathbb{Z}_2^n$ for some $n$.  First let's show that the group is abelian.  Therefore let $g$ and $h$ be group elements.  Because $g$ and $h$ are involutions, we have $(gh)^{-1} = h^{-1} g^{-1} = hg$.  But because $gh$ is also an involution, $(gh)^{-1} = gh$.  So $gh = hg$ as required.  Now the fundamental theorem of finite abelian groups tells us that $G$ must be a direct product of cyclic subgroups of prime-power order.  Because all elements of $G$ are involutions, any cyclic subgroup must be a copy of $\mathbb{Z}_2$, and the result is proven.

Another elementary point to note is that if a coloring automorphism preserves any vertex, then it is necessarily the trivial automorphism that maps all vertices to themselves.  To see this, suppose $\pi(i) = i$ for some vertex $i$, and consider any other vertex $j$.  We have $r_{ij} = r_{\pi(i)\pi(j)} = r_{i\pi(j)}$, and because $r$ is a bijection, for fixed $i$, from vertices $j \neq i$ to colors, we can conclude $\pi(j) = j$. It follows that any permutation that gives a coloring automorphism consists of $q/2$ two-cycles. 

We also note that color automorphisms that have a two-cycle in common must be identical. For suppose there are color automorphisms $\pi$ and $\tilde{\pi}$ and a vertex $i$ such that $\pi(i)=\tilde{\pi}(i)$. Then for any $j$ we have that $r_{ij}=r_{\pi(i)\pi(j)}=r_{\tilde{\pi}(i)\pi(j)}$ but also $r_{ij}=r_{\tilde{\pi}(i)\tilde{\pi}(j)}$. Since $r$ is a bijection, it follows that $\pi(j)=\tilde{\pi}(j)$ for all $j$.

\begin{figure}
\centering{
\includegraphics[height=25ex]{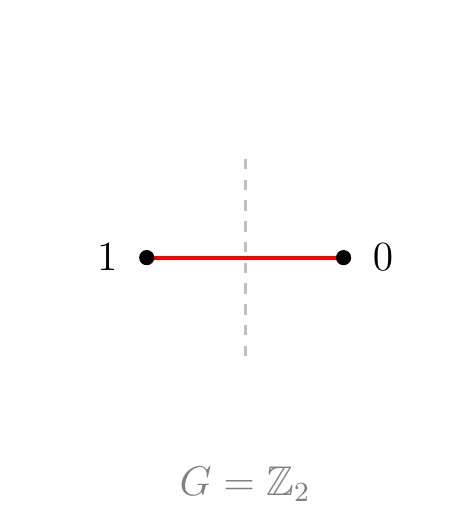}
\includegraphics[height=25ex]{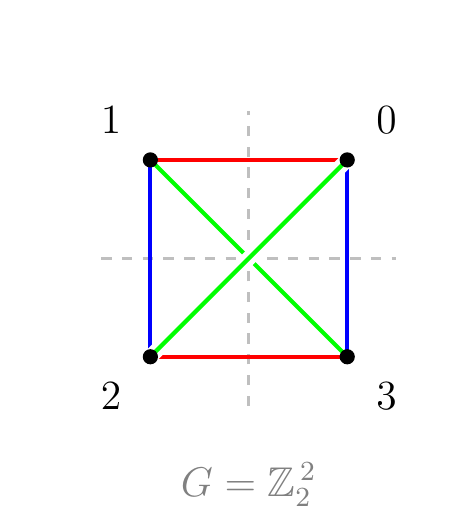}
\hspace{2mm}
\includegraphics[height=25ex]{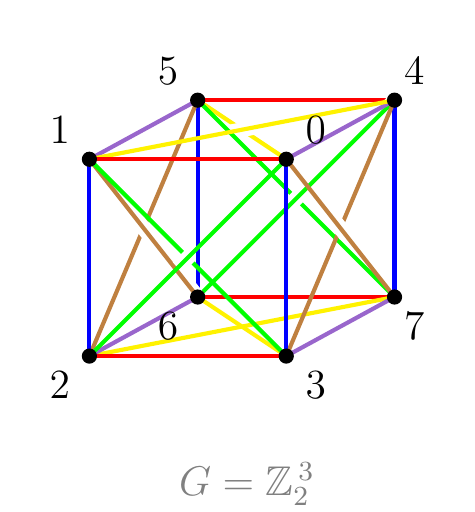}
}
\caption{The first three instances of maximally symmetrical interaction vertices, with $q=2^v$ and $G = \mathbb{Z}_2^v$.  We omitted one color from the $q=8$ case, namely a seventh color with edges running from each corner to the diametrically opposite corner, for example from $0$ to $6$.}
\label{Reflections}
\end{figure}

It helps our imagination to think of the group $G = \mathbb{Z}_2^n$ of coloring automorphisms as reflections through $n$ orthogonal planes which do not pass through any vertices.  If $q=2^v$ and $n=v$, this line of thinking suggests that we can produce a coloring of $K_q$ whose automorphism group is $\mathbb{Z}_2^v$: See figure~\ref{Reflections} for the first few instances.

With these preliminaries in hand, we now come to the main result of this section: If $q=u 2^v$ where $u$ is odd, then the largest that $G$ can be is $\mathbb{Z}_2^v$ if $u=1$, or $\mathbb{Z}_2^{v-1}$ if $u>1$.  The arguments in the remainder of this section do not demonstrate the existence of interaction vertices with any particular symmetry group $G$; rather, they rule out larger symmetry groups.

The proof of our main result relies on the orbit-stabilizer theorem, which we summarize here for the purposes of a self-contained presentation.  If a group $G$ acts on a set $X$, then the stabilizer $\stab i$ of an element $i \in X$ is the subgroup of $G$ of elements which preserve $i$.  Meanwhile, the orbit $\orb i$ is the subset of $X$ consisting of all images of $i$ under the action of elements of $G$.  The theorem says
 \eqn{OrbStab}{
  |\orb i||\stab i| = |G| \,.
 }
As a first application, let $X$ be the set of $q$ vertices, assume that $G = \mathbb{Z}_2^n$.  The stabilizer of a vertex $i$ is the trivial group: This is because any $\pi \in G$ that maps $i$ to itself must also map all other vertices to themselves.  Invoking \eno{OrbStab} we see that each orbit contains $2^n$ points.  The union of all vertex orbits is all of $X$, and so there must be
 \eqn{qtildeDef}{
  \tilde{q} \equiv u 2^{v-n}
 }
distinct vertex orbits.  Already, \eno{qtildeDef} shows that $G$ cannot be larger than $\mathbb{Z}_2^v$, because if $n>v$, $\tilde{q}$ is not an integer.

To finish proving the main result, all we need to do is to exclude the possibility that $G = \mathbb{Z}_2^v$ when $u>1$.  This turns out to require somewhat more subtle reasoning than we have used so far, but the essential idea is to consider the quotient of the interaction vertex by $G$ and show that, as a graph with $\tilde{q}=u$ vertices, it leads to an impossible coloring problem.  Clearly, we could just set $v=n$ throughout the following paragraphs, but we refrain from doing so because keeping $n \leq v$ general allows us to see some first hints on how to actually construct graphs with any allowed symmetry.

Let's start with a second application of the orbit-stabilizer theorem.  Consider the set of all edges {\it of a fixed color}.  For any fixed color, there are $q/2$ such edges.  Edges of a fixed color are permuted among themselves by coloring automorphisms, and for each edge we have
 \eqn{OrbStabAgain}{
  |\orb e||\stab e| = |G| \,.
 }
Following a standard trick, we divide both sides of \eno{OrbStabAgain} by $|\stab e|$ and then sum over all distinct orbits to get
 \eqn{OrbSum}{
  {q \over 2} = \sum_{\rm orbits} {|G| \over |\stab e|} \,.
 }
This can be written more usefully as
 \eqn{OrbSimp}{
  u = 2^{n-v+1} \sum_{\rm orbits} {1 \over |\stab e|} \,.
 }
To get proper mileage out of \eno{OrbSimp}, we need some knowledge of the values of $|\stab e|$---which need not be unity!  Consider however the following division of edges (still of any fixed color) into two classes.  There are edges which join two vertices which are in the same vertex orbit; we will call these ``internal'' edges.  All other edges we will refer to as ``external'' edges.  We observe that $\stab e$ is the trivial group for any external edge, because if it weren't, then there would be some coloring automorphism that exchanges the edge's two ends, and that would make the edge internal.  So $|\stab e| = 1$ for external edges.

Next we want a count of external edges of a fixed color.  It's easier to start by considering external edges of {\it any} color, i.e.~the disjoint union of external edges of each fixed color.  There are $\binom{\tilde{q}}{2} \cdot 2^{2n}$ such edges, because to specify one we must choose a pair of vertex orbits, and then from each of those two vertex orbits we must choose one vertex.  Because the stabilizer of an external edge is trivial, its orbit must have $2^n$ elements.  So the count of external {\it edge orbits} of any color is
 \eqn{Nedge}{
  N_e = \binom{\tilde{q}}{2} \cdot 2^n = q \left( u 2^{v-n-1} - {1 \over 2} \right) \,.
 }
Suppose now $0 \leq n < v$.  Then $2^{v-n-1}$ is a whole number, and because we have only $q-1$ colors to work with, there must be at least one color---call it red---with at least $u 2^{v-n-1}$ external edge orbits.  Restricting to red edges only, we recall that $|\stab e| = 1$ for each external edge, and so at least $u 2^{v-n-1}$ terms in the sum on the right hand side of \eno{OrbSimp} must be unity.  Comparing to \eno{OrbSimp}, we see that---for red edges---the sum over orbits works out perfectly with only external edges, implying that there can't be any red internal edges.  This is informative and useful for constructing examples.

\begin{figure}
  \centerline{\includegraphics[width=5in]{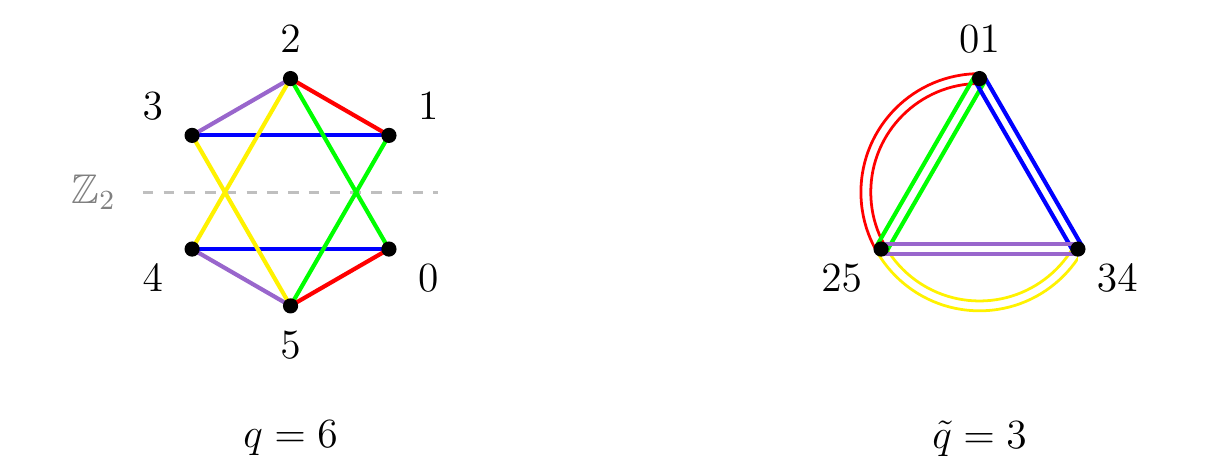}}
  \caption{A partial coloring for $q=6$ which illustrates what goes wrong when one demands too much symmetry.  Here we propose $\mathbb{Z}_2$ symmetry, which according to our general result is too much.  On the left we show a partial coloring with $\mathbb{Z}_2$ symmetry.  On the right, we show the same coloring modded out by the $\mathbb{Z}_2$ symmetry.  The dots on the right are vertex orbits, and the double lines of each color are edge orbits.  The problem is that we need two edge orbits between each pair of vertex orbits, and all of them (in this simple case) must have different colors.  That means we need six colors just for the external edges, and we only have five to work with---leaving us without a second color to use in connecting the $01$ and $34$ orbits.  Internal edges would run within a vertex orbit, for example from $0$ to $1$.}\label{SymmetryFailure}
 \end{figure}

Now suppose $n=v$.  If also $u=1$, then from \eno{Nedge} we see that $N_e=0$: There are no external edges at all!  This makes sense because there is only one vertex orbit, and indeed the cubical vertex illustrated in figure~\ref{Reflections} shows that it is entirely consistent to have $n=v$ and $u=1$.  Where things get dangerous is if $n=v$ and $u>1$.  Then, from \eno{Nedge}, $N_e = q(u-1)/2$.  Since we have only $q-1$ colors to work with, there must be at least one color---again call it red---with at least $(u+1)/2$ external edge orbits.  Considering only red edges the sum on the right hand side of \eno{OrbSimp} restricted to external edges gives $u+1$.  This is disastrous, because adding in the contribution of internal edges (if any) results in the absurd inequality $u \geq u+1$.  Another way to put it is that if $n=v$ and $u$ is odd and greater than $1$, then we can't color even the external edges consistently with only $q-1$ colors---let alone the internal edges.  See figure~\ref{SymmetryFailure}.

\section{Construction of interaction vertices}
\label{CONSTRUCTION}

Since for any $q$ we can construct an interaction vertex with the canonical ordering, which has no coloring automorphisms, what remains to be shown is that for $q$ equal to twice times an even number, it is always possible to construct an interaction vertex with any allowed non-trivial symmetry group. Figure~\ref{Octagons} explicitly shows that this is true for $q=8$. We now prove inductively that it is also true for $q>8$.
 \begin{figure}
  \centering

\begin{subfigure}[b]{0.3\textwidth}
\includegraphics[width=\textwidth]{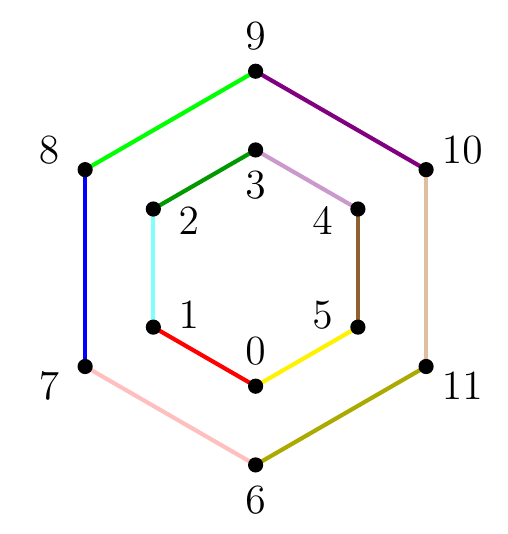}
        \caption{}
        \label{ThreeHexagonsA}
\end{subfigure}
\hspace{5mm}
  \begin{subfigure}[b]{0.3\textwidth}
\includegraphics[width=\textwidth]{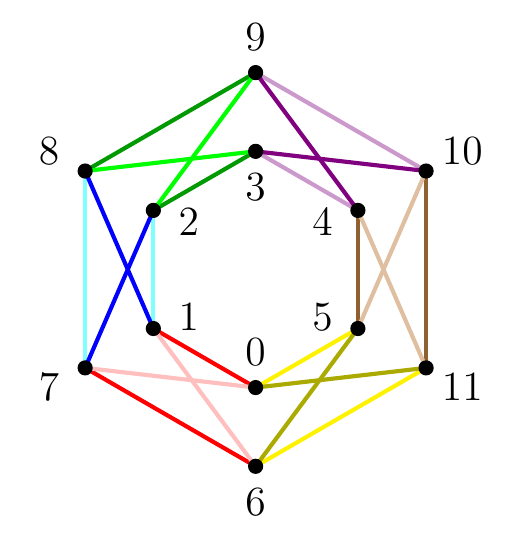}
        \caption{}
        \label{ThreeHexagonsB}
\end{subfigure}
\hspace{5mm}
\begin{subfigure}[b]{0.3\textwidth}
\includegraphics[width=\textwidth]{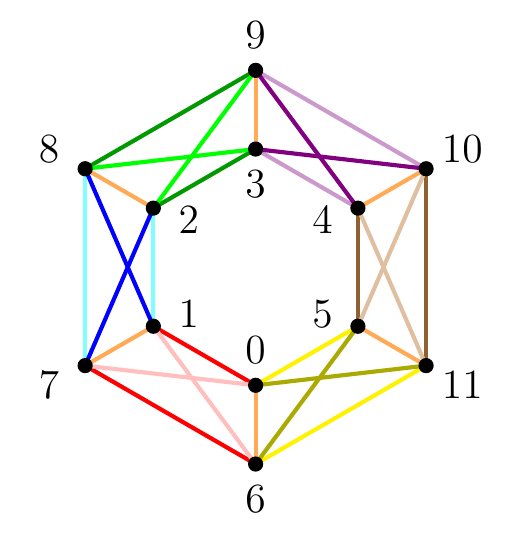}
        \caption{}
        \label{ThreeHexagonsC}
\end{subfigure}

  \caption{Constructing an interaction vertex with $q=12$ and $G = \mathbb{Z}_2$ in three steps.  (a) Separate the vertices into two groups of $6$, shown here as inner and outer rings.  We've connected each group only cyclically in order to avoid clutter, but in the full construction, we start with two copies of the complete graph $K_6$, one colored according to $r_{ij}$ and the other according to $\tilde{r}_{ij} = r_{ij} + q-1$.  (b) Reconnect the $12$ vertices according to the prescription shown in the first two lines of \eno{InductiveColoring}.  (c) Add the last color according to the third line of \eno{InductiveColoring}.  In the parlance of section~\ref{SYMMETRY}, these last edges are the internal edges, while the ones colored in the previous step are the external edges.}\label{ThreeSteps}
 \end{figure}

Let $q'=q/2$ and assume that $r_{ij}$ is a coloring of $K_{q'}$ with symmetry group $G'$ (which may be the trivial group). Split the $q$ vertices into two sets of $q'$ vertices; notationally this can be done by identifying a vertex first by indicating which set it's in, say with a Greek index $\alpha=0$ or $1$, and then which vertex within the set it is, say with a Roman index $i \in \{0,1,\dots,q-1\}$. Let $\tilde{r}_{ij}\equiv r_{ij} + q-1$. We now claim that the coloring $r_{\alpha i,\beta j}$ of $K_q$ defined by
\eqn{InductiveColoring}
{
& r_{0i,0j}=r_{1i,1j}=r_{ij}\,, \cr
& r_{0i,1j}=r_{1i,0j}=\tilde{r}_{ij}\,, \cr
& \tilde{r}_{0i,1i}=\tilde{r}_{1i,0i}=q-1\,,
}
has symmetry group $G=G'\times \mathbb{Z}_2$. See figure~\ref{ThreeSteps} for a diagrammatic illustration of how this coloring is generated. That the coloring \eqref{InductiveColoring} inherits the $G'$ symmetry is clear from the fact that, for any $\pi \in G'$, we have that
\eqn{}{
r_{\alpha \pi(i),\beta \pi(j)}=r_{\alpha i,\beta j}\,. }
But by construction the coloring \eqref{InductiveColoring} is also invariant under the permutation $\tau$ that acts on the Greek indices as 
\eqn{}{\alpha \rightarrow \alpha +1 \mod{2}\,, \hspace{10mm} \beta \rightarrow \beta +1 \mod{2}\,,}
while leaving the Roman indices unchanged. Furthermore, the coloring \eqref{InductiveColoring} has no other symmetries. For coloring automorphisms that only swap around Roman indices are in one-to-one correspondence with the coloring automorphisms of $r_{ij}$. And if a coloring automorphism $\sigma$ changes the Greek index of any index pair $\alpha i$ such that $\sigma(0i)=1i'$, then for any $j$
\eqn{}
{
r_{\sigma(0i),\sigma(0j)}=r_{1i',\sigma(0j)}\,.
}
But since $r_{0j,0k} \neq r_{1j',0k'}$ for all $j,j',k,k'$, it follows that $\sigma$ must change the Greek index of \emph{all} index pairs and so must be of the form $\sigma = \pi \circ \tau$ for some $\pi \in G'$. This completes the inductive proof.

In general, for a given symmetry, there are multiple interaction vertices different from the one generated by the prescription \eqref{InductiveColoring}. An exception, however, occurs for the maximally symmetric vertex when $q=2^v$. In this case there is only one vertex with $G=\mathbb{Z}_2^{\,v}$. For $G$ consists of the identity element and $q-1$ permutations that commute amongst each other and each consists of $q/2$ two-cycles. And amongst permutations in $S_{q}$ consisting of $q/2$ two-cycles, one can at most form a set of $q-1$ elements that commute amongst each other but don't share a two-cycle; and any two such sets are equivalent by conjugation. But if we consider the edges of a given color, say red, in a maximally symmetric colored graph, then this sub-graph is invariant under the permutation $\pi$ that swaps vertices connected by a red edge. But $\pi$ must commute with all permutations in $G$, and so it follows that $\pi \in G$. Moreover, if we explicitly write $\pi = (a_0a_1)(a_2a_3)\dots(a_{q-1}a_q)$, then we recognize that each two-cycle in $\pi$ corresponds to a red edge; in other words, each non-trivial element of $G$ is precisely associated with a one-factor.  Since there are $q-1$ non-trivial elements of $G$, the one-factors are all specified once $G$ is specified.  And we have argued that $G$ is essentially unique.

\section{One-factorizations and equivalent interaction terms}
\label{ONEFACTOR}
In this section we determine the conditions under which two theories with actions of the form (\ref{SKTextended}) are equivalent. As discussed briefly in section \ref{STRUCTURE}, the choice of an interaction tensor $\Omega_{A^{(0)}A^{(1)}\cdots A^{(q-1)}}$ corresponds to a coloring problem on the complete graph $K_q$ on $q$ vertices. 

More precisely, these interaction tensors are in one-to-one correspondence with ordered one-factorizations of $K_q$.  A one-factor of $K_q$ is a set $F$ of edges such that each vertex of $K_q$ belongs to a unique edge in $F$.  A one-factorization of $K_q$ is a partition of the edge-set of $K_q$ into $q-1$ one-factors $F_i$.  We denote a one-factorization as $\overline{\mathcal{F}} = \{F_0,\dots,F_{q-2}\}$, and when speaking of one-factorizations we do not distinguish between different orderings of the $F_i$.  An ordered one-factorization is a $(q-1)$-tuple $\mathcal{F} = (F_0,\dots,F_{q-2})$, i.e.~we have imposed a particular order on a one-factorization.  To see that ordered one-factorizations correspond to interaction vertices, recall that the fields $\psi^{a_0\ldots a_{q-2}}$ don't have any built-in symmetry under interchange of indices $a_i$.  Each such index is associated with a definite color: For example, we could say that red is associated with the first index $a_0$, then green with $a_1$, blue with $a_2$, and so forth.  And each color is associated with a one-factor: red for $F_0$, green for $F_1$, blue for $F_2$, etc. 

One can check that any graph automorphism $\pi\colon K_q\to K_q$ carries (ordered) one-factori\-zations into other (ordered) one-factorizations via the actions $\pi \mathcal{F}=\left(\pi(F_0),\ldots,\pi(F_{q-2})\right)$ and $\pi \overline{F} \equiv \overline{\pi\mathcal{F}}$, respectively. Two ordered one-factorizations $\mathcal{F}$ and $\mathcal{G}$ are said to be isomorphic if there exists a graph automorphism $\pi$ such that $\mathcal{F}=\pi\mathcal{G}$. Similarly, $\overline{\mathcal{F}}\simeq\overline{\mathcal{G}}$ means there exists $\pi$ so that $\overline{\mathcal{F}}=\pi\overline{\mathcal{G}}$. We can also consider the natural action of any $\tau\in S_{q-1}$ on ordered one-factorizations $\tau\mathcal{F}=(F_{\tau(0)},\ldots,F_{\tau(q-2)})$. This action simply permutes the colors associated to each one-factor. In this language, $\overline{\mathcal{F}}\simeq\overline{\mathcal{G}}$ if and only if there exists a graph automorphism $\pi$ and a permutation of colors $\tau$ such that $\pi \mathcal{F}=\tau \mathcal{G}$.

Given an ordered one-factorization $\mathcal{F}$, we can form the following interaction tensor:

\eqn{OFTensor}{
\Omega^\mathcal{F}_{A^{(0)}\ldots A^{(q-1)}} \equiv \prod_{r=0}^{q-2}\prod_{\langle i j \rangle \in F_r}\Omega_{a^{(i)}_r a^{(j)}_r} \, ,
}
where $\langle i j \rangle$ is the edge in $K_q$ running between the vertices $i$ and $j$ and $a_r$ is the color index associated to the one-factor $F_r$. In writing $\langle i j \rangle$, we implicitly assume $i<j$. Conversely, any degree $q$ interaction term in which each constituent $\psi^A$ has a single color index contracted with each other $\psi^A$ arises from an interaction tensor of the form (\ref{OFTensor}).

Note that for distinct ordered one-factorizations $\mathcal{F}$ and $\mathcal{G}$ we end up with distinct interaction tensors $\Omega^\mathcal{F}\neq\Omega^\mathcal{G}$. We will however show that if their underlying one-factorizations $\overline{\mathcal{F}}$ and $\overline{\mathcal{G}}$ are isomorphic then $\Omega^\mathcal{F}$ and $\Omega^\mathcal{G}$ give rise to equivalent theories. Our argument proceeds in two parts: First we discuss equivalence under the action of $\tau \in S_{q-1}$ permuting colors, and then we consider the action of a permutation $\pi \in S_q$ of vertices.

Suppose we make the following field redefinition: $\phi^A=\psi^{\tau^{-1} A}$, where $\tau^{-1}\in S_{q-1}$ and $\tau A \equiv a_{\tau^{-1}(0)}\ldots a_{\tau^{-1}(q-2)}$. Then:
\eqn{OFColorEquiv}{
\Omega^\mathcal{F}_{A^{(0)}\ldots A^{(q-1)}}\phi^{A^{(0)}}\cdots\phi^{A^{(q-1)}}=\Omega^{\tau \mathcal{F}}_{A^{(0)}\ldots A^{(q-1)}}\psi^{A^{(0)}}\cdots\psi^{A^{(q-1)}}\,.
}
So, modulo a linear field redefinition implemented by a permutation matrix, the interaction term only depends on the underlying one-factorization $\overline{\mathcal{F}}$.

Now view a graph automorphism as a permutation on $q$ vertices, $\pi\in S_q$. With the change of indices $B^{(i)}=A^{\pi(i)}$ one can show:
\eqn{OGIsoEquiv}{
(\sigma_\psi)^\pi \Omega^\mathcal{F}_{A^{(0)}\ldots A^{(q-1)}}\psi^{A^{(0)}}\cdots\psi^{A^{(q-1)}}&= \Omega^\mathcal{F}_{A^{(0)}\ldots A^{(q-1)}}\psi^{A^{\pi^{-1}(0)}}\cdots\psi^{A^{\pi^{-1}(q-1)}}
\cr
 &=\Omega^{\pi \mathcal{F}}_{B^{(0)}\ldots B^{(q-1)}}\psi^{B^{(0)}}\cdots\psi^{B^{(q-1)}}\,.
}
So any two isomorphic ordered one-factorizations give the same interaction term, up to a sign.\footnote{In the presence of multiple interaction terms in the lagrangian, this argument that a permutation of vertices does not lead to a new theory continues to hold true. But it will no longer be true that an interaction term only depends on the underlying one-factorization $\overline{\mathcal{F}}$, see appendix \ref{IsomorphismOrdered}. }
We can get rid of this sign by sending $g\to -g$.\footnote{See \cite{Bulycheva:2017ilt} and \cite{Klebanov:2018nfp} for a careful discussion, in the context of the $q=4$ fermionic tensor model, of how the Hamiltonian even at the quantum level transforms in the degree 1 sign representation under permutations of indices and of how this affects the spectrum of the theory.}

We should point out here that isomorphic one-factorizations give equivalent interactions to all orders in $N$. As we will show in sections~\ref{TWOPOINT} and~\ref{FOURPOINT}, even non-isomorphic one-factorizations lead to the same two-point and four-point functions in the melonic limit, up to a rescaling of $g$ by a power of the order of the coloring automorphism group $G$.
\begin{table}
\centering
\begin{tabular}{|c|c|c|c|c|c|c|c|}\hline
\diagbox[innerwidth=6 ex,height=5ex]{$G$}{$q$}& $2$ & $4$ & $6$ & $8$ & $10$ & $12$ & $14$\\\hline
$1$                          &  0 &  0 & 1  &  2 & 396   &   526,910,769 & $\sim 1.13 \times 10^{18}$\\\hline
$\mathbb{Z}_2$     &  1 &  0 &  0 &  1 &  0  &   4851 & 0 \\\hline
$\mathbb{Z}^2_2$ & 0 &  1 &  0 & 2  &  0 &   0 & 0 \\\hline
$\mathbb{Z}^3_2$ & 0 &  0 &  0 &  1 &  0 &   0 & 0 \\\hline
\end{tabular}
\caption{The number of isomorphism classes of one-factorizations of $K_q$ with given symmetry group $G=\mathrm{Aut}(\mathcal{F})$. The values for $q=12$ and $14$ were determined using  \cite{Dinitz:1994} and \cite{Kaski:2009}, respectively; the exact count for $q=14$ is quoted in the main text.}
\label{OFTable}
\end{table}

For future reference, we note that by $\mathrm{Aut}(\mathcal{F})$ we mean all graph automorphisms $\pi$ such that $\pi \mathcal{F}=\mathcal{F}$. $\mathrm{Aut}(\mathcal{F})$ is the exactly the vertex automorphism group described in section \ref{SYMMETRY}. In contrast, $\mathrm{Aut}(\overline{\mathcal{F}})$ consists of all graph automorphisms $\pi$ such that $\pi \overline{\mathcal{F}}=\overline{\mathcal{F}}$, that is all $\pi$ that can be undone by a permutation $\tau\in S_{q-1}$ of the colors $\pi \mathcal{F}=\tau \mathcal{F}$. Clearly $\mathrm{Aut}(\mathcal{F})\leq \mathrm{Aut}(\overline{\mathcal{F}})$. In general $\mathrm{Aut}(\overline{\mathcal{F}})$ has a much richer group structure than $\mathrm{Aut}(\mathcal{F})$. See for instance table~\ref{AutTable} in appendix~\ref{AlgorithmAppendix} where, for $q=12$, we list the ten possible values for $|\mathrm{Aut}(\overline{\mathcal{F}})|$ when $|\mathrm{Aut}(\mathcal{F})|=2$.

 In the case $q=12$, the vast majority of isomorphism classes of one-factorizations have trivial $\mathrm{Aut}(\mathcal{F})$. It is known in the literature \cite{Dinitz:1994} that there are a total of 526,915,620 non-isomorphic one-factorizations (the sum of the two entries in the $q=12$ column in figure \ref{OFTable}) and exactly 252,282,619,805,368,320 distinct one-factorizations of $K_{12}$. Implementing the orderly algorithm used in \cite{Seah} and \cite{Dinitz:1994} on the set of one-factors invariant under a given involution $\pi \in S_{12}$, we find that there are 1,008,649,635,840 one-factorizations of $K_{12}$ with $\mathrm{Aut}(\mathcal{F})=\mathbb{Z}_2$ and that they fall into 4851 isomorphism classes under permutations of vertices, see appendix \ref{AlgorithmAppendix}.
 
For $q=14$, $\mathrm{Aut}(\mathcal{F})$ can only be trivial and there are 1,132,835,421,602,062,347 nonisomorphic one-factorizations of $K_{14}$ \cite{Kaski:2009}. 

For $q \leq 14$, a summary of the number of isomorphism classes of one-factorizations and therefore the number of inequivalent interaction terms is given in table~\ref{OFTable}. It is known that for sufficiently large $q$, the number of non-isomorphic one-factorizations $N(q)$ satisfies $\log N(q) \sim \frac{1}{2}q^2 \log q$ \cite{Cameron:1976}. In particular, $N(q)\to\infty$ as $q\to \infty$.

\section{The two-point function and the Schwinger-Dyson equation}
\label{TWOPOINT}

The kinetic term in the action \eqref{SKTextended} gives rise to a free propagator that in momentum space is given by
\eqn{G0momentum}
{
G_0^{a_0a_1...a_{q-2}b_0b_1...b_{q-2}}(\omega)=\sqrt{\sgn(-1)}\,\frac{\sgn(\omega)}{|\omega|^s}\prod_{i=0}^{q-2}\Omega^{a_ib_i}\equiv G_0(\omega)\prod_{i=0}^{q-2}\Omega^{a_ib_i}\, ,
}
where the matrix $\Omega^{ab}$ is the inverse of $\Omega_{ab}$, that is, $\Omega_{ab}\Omega^{bc}=\delta_a^c$.  With the normalization of the kinetic term as given in \eno{SKTextended}, we must choose
 \eqn{sqrtSgn}{
  \sqrt{\sgn(-1)} = \left\{ \seqalign{\span\TL &\quad \span\TT}{
    1 & if $\sgn(-1) = 1$  \cr
    i & if $\sgn(-1) = -1$\,.} \right.
 }

\begin{figure}[b]
 \centerline{\includegraphics[width=4.7in]{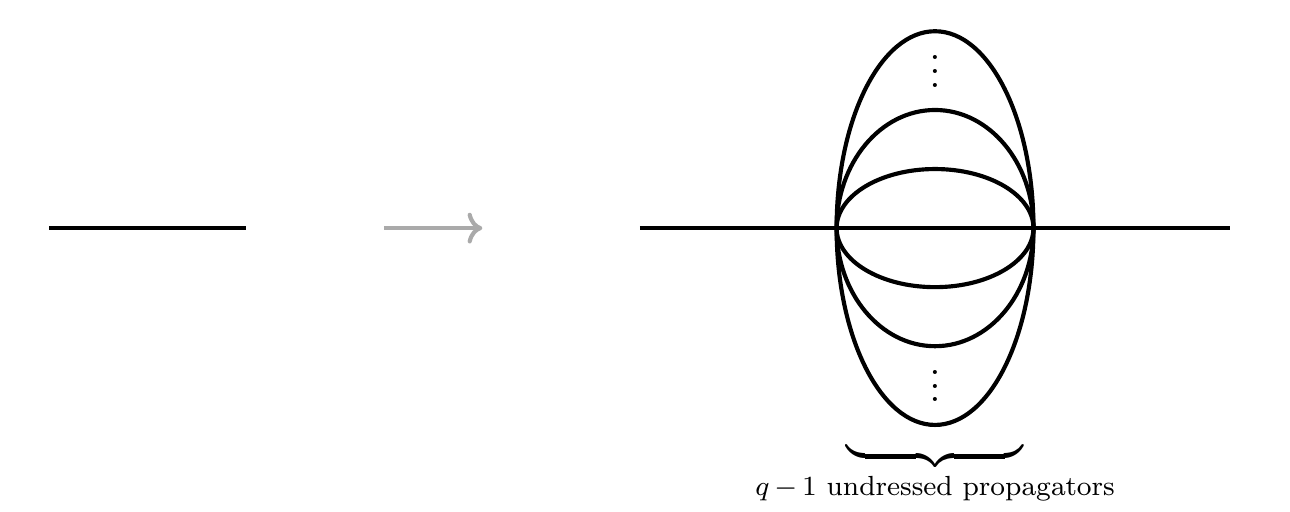}}
 \caption{A melonic insertion.}\label{GeneralInsertion}
\end{figure}

Once we include an interaction term, the propagator picks up loop corrections, but in the limit where $g\rightarrow 0$ and $N\rightarrow \infty$ such that $g^2N^{\frac{(q-1)(q-2)}{2}}$ is kept fixed, only melonic diagrams survive. The melonic contributions to the free propagator can all be obtained by iteratively applying the melonic insertion to the free propagator as shown on figure~\ref{GeneralInsertion}. 

\begin{figure}[t]
 \centerline{\includegraphics[width=5in]{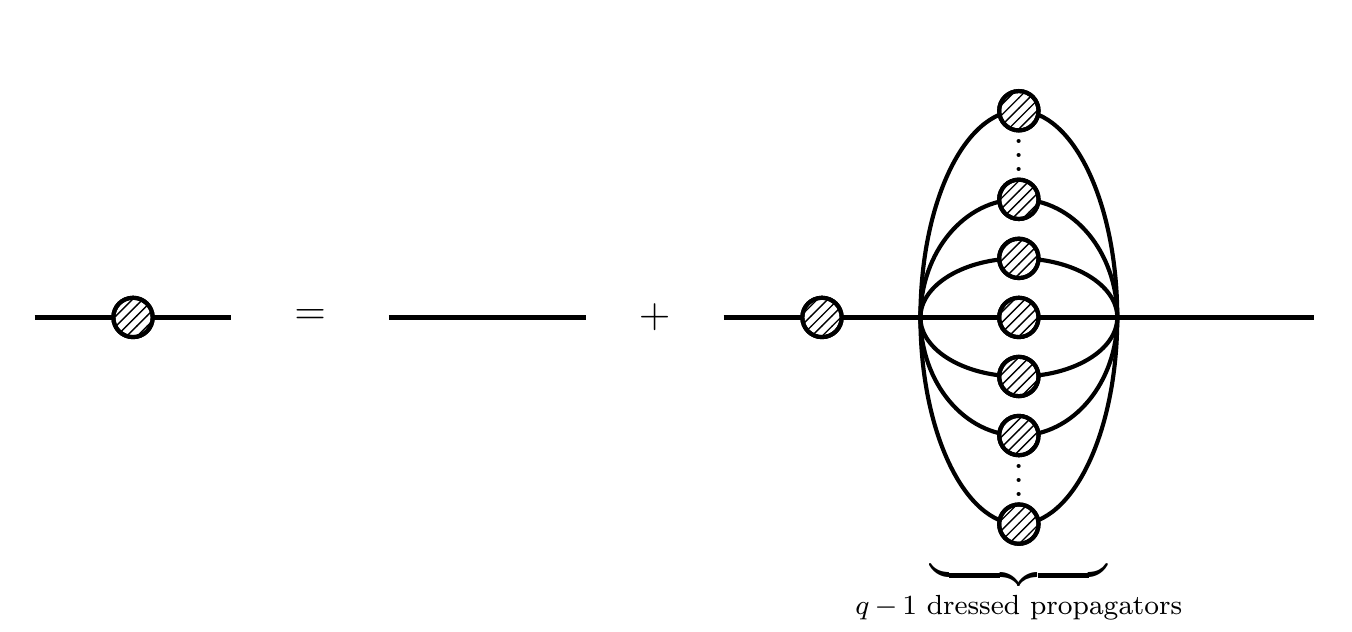}}
 \caption{The Schwinger-Dyson equation for the dressed propagator.}\label{SchwingerDyson}
\end{figure}

Adding together the free propagator and all the melonic corrections yields the dressed propagator
\eqn{}
{
G^{a_0a_1...a_{q-2}b_0b_1...b_{q-2}}(\omega) \equiv G(\omega)\prod_{i=0}^{q-2}\Omega^{a_ib_i}\,.
}
Just as in the cases of the SYK model \cite{Polchinski:2016xgd} and the Klebanov-Tarnopolsky tensor model with $q=4$ \cite{Klebanov:2016xxf}, the dressed propagator satisfies the Schwinger-Dyson equation depicted schematically in figure~\ref{SchwingerDyson}.  It first seems surprising that the right-most propagator in figure~\ref{SchwingerDyson} has no melonic insertion.  This is correct because any Feynman diagram contributing to the dressed propagator must have a right-most melonic insertion, and a free propagator attaches to it from the right.

To work out the exact mathematical expression for the Schwinger-Dyson equation, it is necessary to consider the automorphism group of the interaction as well as the number of sign flips involved in index contraction. 

\begin{figure}[b]
 \centerline{\includegraphics[width=5in]{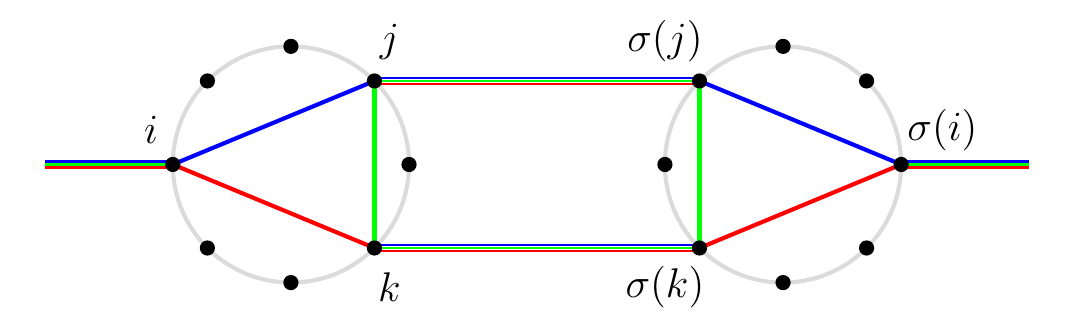}}
 \caption{To get a melonic diagram, the external legs must be connected to vertices in the two interactions that are related by an automorphism.}\label{2ptPermutation}
\end{figure}

The first melonic correction to the free propagator, depicted on the right-hand side of figure~\ref{GeneralInsertion}, contains two interactions each of which is described by a colored graph with vertices labeled from 0 to $q-1$. For a given Feynman diagram corresponding to the first melonic correction, define a permutation $\sigma \in S_q$ by requiring that if an external propagator enters the left interaction vertex at vertex $i$, then an external propagator enters the right interaction vertex at vertex $\sigma(i)$; and likewise requiring that an internal propagator connected to vertex $j$ in the left interaction vertex connects to vertex $\sigma(j)$ in the right interaction vertex. See figure~\ref{2ptPermutation}.

We claim now that such a Feynman diagram is melonic if and only if $\sigma$ belongs to the automorphism group $G$ of the interaction. The claim can be proved by the following reasoning:
\begin{itemize}
\item The two interactions are connected by $q-1$ internal propagators, each of which carry $q-1$ threads, giving a total of $(q-1)^2$ internal threads.
\item The $q-1$ threads of the left external propagator must connect to the $q-1$ threads of the right external propagator, so there are at most $(q-1)(q-2)$ internal threads that can partake in index loops.
\item Since at least two internal threads are needed for an index loop, there can be at most $(q-1)(q-2)/2$ index loops---the number required for a melonic diagram. This maximum number of loops is achieved when and only when the external threads only pass once through the internal propagators and the each index loop contains exactly two internal threads.
\item Labeling the vertex that the left external propagator is incident to by $i$, the fact that the external threads must pass only once through the internal propagators is equivalent to the statement that $r_{ij}=r_{\sigma(i)\sigma(j)}$ for all $j\neq i$.
\item The fact that each index loop must contain exactly two internal threads is equivalent to the statement that for any two indices $j$ and $k$ different from $i$, 
$r_{jk}=r_{\sigma(j)\sigma(k)}$.
\end{itemize}
We see then that the melonic diagrams are exactly those for which $r_{lm}=r_{\sigma(l)\sigma(m)}$ for all $l,m$, ie. those for which $\sigma \in G$. Since in writing down the Feynman diagram corresponding to the first melonic correction we may take the left external propagator to be incident to any vertex $i$, while the right external propagator can only be incident to vertices $i'$ for which $i'=\sigma(i)$ for some $\sigma \in G$, the total number of Feynman diagrams contributing is $q|G|$. And for each additional melonic insertion we pick up an extra copy of this factor.

\begin{figure}
 \centerline{\includegraphics[width=5in]{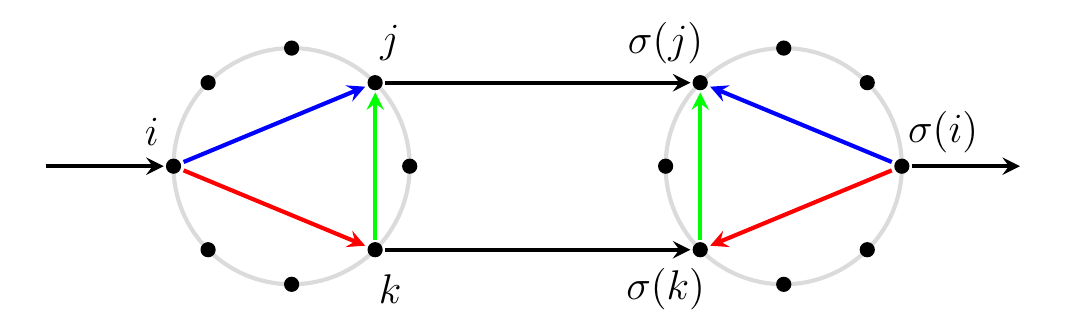}}
 \caption{All index loops can be brought to have a uniform orientation by an even number of arrow flips. To get a uniform orientation of any external thread an odd number of arrows flips are requisite.}\label{2ptArrows}
\end{figure}

The other subtlety to consider before writing down the Schwinger-Dyson equation is whether the contraction of index loops causes a sign difference between $O(N)$ and $Sp(N)$ symmetric models. To account diagrammatically for the possibility of $Sp(N)$ symmetry, we replace each thread with an arrow. The orientation of any arrow can be flipped at the cost of a factor of $\sigma_\omega$, and index loops of arrows with uniform orientation can be contracted to give a factor of $N$ without picking up a sign.

For the first melonic correction to the free propagator we may take the threads of all propagators to point from left to right as in figure~\ref{2ptArrows}. For a given Feynman diagram described by a permutation $\sigma \in G$ as explained above, we may for any $l$ and $m$ take the the arrow between vertices $l$ and $m$ to have the same orientation as the arrow between vertices $\sigma(l)$ and $\sigma(m)$ since any automorphism involves an even number of arrow flips. In this case, all index loops consist of two arrows with one orientation and two arrows with the opposite orientation, and so all the loops may be contracted without picking up a sign. But to obtain a uniform orientation of the external threads running from left to right and recover the index structure of the free propagator, it is necessary for each of the $q-1$ colors to flip one thread in one of the two interactions. Hence, we must in total flip an odd number off arrows, and so we conclude that we pick up a factor of $\sigma_\Omega$ with each melonic insertion.  This fact was shown in \cite{Gubser:2017qed} to apply for $q=4$, but we see now that it holds true for all $q$.

Taking into account the symmetry of the interaction and the index structure of the corrections, we arrive finally at the Schwinger-Dyson equation
\eqn{eqSchwingerDyson}{
G(t_1-t_2) &= G_0(t_1-t_2)  \cr
 &\qquad{} +
q|G|\left(\frac{g}{|G|}\right)^2\sigma_\Omega\, N^{\frac{(q-1)(q-2)}{2}}\int dt\,dt'\,G(t_1-t)G(t-t')^{q-1}G_0(t'-t_2)\, .
}
To slightly shorten expressions and remove the explicit dependence on $G$, we find it useful to introduce the definition 
\eqn{gtilde}
{
\tilde{g} \equiv \sqrt{\frac{q}{|G|}N^{\frac{(q-1)(q-2)}{2}}}\,g \,.
}

\subsection{The IR solution}
\label{sec:IRsolution}

The Schwinger-Dyson equation can be expressed in term of convolutions as
\eqn{SchwingerDysonConvolution}
{
G(t)=G_0(t) + \sigma_\Omega \tilde{g}^2 (G\ast G^{q-1} \ast G_0)(t)\,.
}
In the infra-red limit, $\tilde{g}$ is large, and to leading order in $1/\tilde{g}$ we can set $G(t)=0$. To solve \eqref{SchwingerDysonConvolution} in the IR, we will adopt the methodology of \cite{Gubser:2017qed}, and so we introduce multiplicative characters defined as
\eqn{multiplicativecharacters}
{
\pi_s(t) \equiv |t|^s, \hspace{10mm} \pi_{s,\sgn}(t) \equiv |t|^s\sgn(t)\,.
}
The Fourier transform of these multiplicative characters are given by 
\eqn{Fourier}
{
\mathcal{F}[\pi_s](\omega)=\Gamma(\pi_{s+1})\,\pi_{-s-1}(\omega)\, , \hspace{10mm}
\mathcal{F}[\pi_{s,\sgn}](\omega)=\Gamma(\pi_{s+1,\sgn})\,\pi_{-s-1,\sgn}(\omega) \, ,
}
where $\Gamma(\pi_{s})$ and $\Gamma(\pi_{s,\sgn})$ are instances of Gel'fand-Graev gamma functions, see appendix \ref{GammaAppendix}. Using the above, we may express the bare propagator in momentum and position space as
\eqn{}
{
G_0(\omega)=\sqrt{\sgn(-1)}\,\pi_{-s,\sgn}(\omega)\, , \hspace{10mm} G_0(t)=\sqrt{\sgn(-1)}\,\Gamma(\pi_{1-s,\sgn})\,\pi_{s-1,\sgn}(t) \,.
}

As an ansatz for solving \eqref{SchwingerDysonConvolution} with the left-hand side set to zero, we choose
\eqn{ansatz}
{
G(t) = b\,\pi_{-\frac{2}{q},\sgn}(t)\, ,
}
and so the IR Schwinger-Dyson equation in position space reads
\eqn{}
{
-\pi_{1-s,\sgn}=\sigma_{\Omega}\,\tilde{g}^2\,b^q\,\pi_{-\frac{2}{q},\sgn}\ast\pi_{\frac{2(1-q)}{q},\sgn}\ast\pi_{1-s,\sgn}\, .
}
Using \eqref{Fourier} to Fourier transform the multiplicative characters, we get the momentum space Schwinger-Dyson equation
\eqn{}
{
-\pi_{s,\sgn}=\sigma_{\Omega}\,\tilde{g}^2\,b^q\,\Gamma(\pi_{\frac{q-2}{q},\sgn})\Gamma(\pi_{\frac{2-q}{q},\sgn})\,\pi_{\frac{2-q}{q},\sgn}\ast\pi_{\frac{(q-2)}{q},\sgn}\ast\pi_{s,\sgn}\,.
}
The multiplicative characters all cancel, and we read off directly that
\eqn{bsol}
{
b^q = \frac{-1}{\sigma_\Omega\, \tilde{g}^2\, \Gamma(\pi_{\frac{q-2}{q},\sgn})\Gamma(\pi_{\frac{2-q}{q},\sgn})}\,.
}
The $s$ dependence is seen to cancel out entirely. Insofar as the ansatz \eqref{ansatz} describes the true IR behavior, the dressed propagators of the various non-local theories with different values of $s$ all flow to the same function in the IR. In the case of bosonic theories, however, where the potential can be unbounded from below, the physical significance of the IR solution \eno{ansatz} is uncertain. Nonetheless, we explicitly include bosonic theories under the scope of theories we are subjecting to formal perturbation theory since bosonic theories are required in order to write down an adelic relation connecting Archimedean and $p$-adic theories, as we will see in section \eqref{ADELIC}. 

In section \ref{ExactSolution} we will see that for a large subset of the $p$-adic theories, the full Schwinger-Dyson equation \eqref{eqSchwingerDyson} can be solved exactly, and in these cases one can verify by inspection of the full solution that the two-point function flows to the IR solution described by equations \eqref{ansatz} and \eqref{bsol}.

\subsection{The zoo of theories}

As we are working in Euclidean space, the reality of the action \eqref{SKTextended} dictates that $G(t)$ should be real too. The ansatz \eqref{ansatz} can therefore only provide an accurate description of the IR behavior if $b$ is real, that is, if the right-hand side of \eqref{bsol} is positive. In other words, the sign of $\Gamma(\pi_{\frac{q-2}{q},\sgn})\Gamma(\pi_{\frac{2-q}{q},\sgn})$ must be opposite to that of $\sigma_\Omega$. For any choice of number field $\mathbb{R}$ or $\mathbb{Q}_p$ and any choice of sign function in the action, this requirement together with the sign constraint \eqref{sgnConstraint} uniquely specifies whether the theory must be bosonic or fermionic and $O(N)$ and $Sp(N)$ symmetric. Generalizing the table of theories in \cite{Gubser:2017qed} to all values of $q$, table~\ref{ExplicitResults} lists the melonic theories with renormalization group flow from a free theory in the UV to a strongly interacting fixed point in the IR.  The choice of $\tau$ in table~\ref{ExplicitResults} amounts to a choice of the sign function, as explained in appendix~\ref{GammaAppendix}.  Often, the value of $\tau$ involves $\epsilon$, which stands for any integer that is not a square modulo $p$.

\definecolor{Gray}{gray}{0.9}
\def\gr{\rowcolor{Gray}}
\renewcommand{\arraystretch}{1.3}
 \begin{table}\begin{center}\begin{tabular}{|c||c|r|c|r|r|l|}
  \hline
  $K$ & condition & $\tau$ & 
    $\Gamma(\pi_{\frac{q-2}{q},\sgn})\Gamma(\pi_{\frac{2-q}{q},\sgn})$ &
    $\sigma_\Omega$ & $\sigma_\psi$ & explanation \\[5pt]
    \hline\hline 
  $\mathbb{R}$ & & $1$ & $-\frac{2\pi q}{q-2}\tan\left(\frac{\pi}{q}\right)$ & $1$ & $1$ & ${\rm O}(N)$ bosonic \\ [5pt]
  $\mathbb{R}$ & & $-1$ & $-\frac{2\pi q}{q-2}\cot\left(\frac{\pi}{q}\right)$ & $1$ & $-1$ & ${\rm O}(N)$ fermionic   \\ [5pt]\hline  
  $\mathbb{C}$ & & $1$ & $-\frac{\pi^2q^2}{(q-2)^2}$ & $1$ & $1$ & ${\rm O}(N)$ bosonic  \\ [5pt] \hline
  $\mathbb{Q}_p$ & $p$ odd & $1$ & $-\frac{(p^{2/q}-1)(p^2-p^{2/q})}{p(p-p^{2/q})^2}$ & $1$ & $1$ & ${\rm O}(N)$ bosonic  \\ [5pt]
  $\mathbb{Q}_p$ & $p$ odd & $\epsilon$ & $\frac{(p^{2/q}+1)(p^2+p^{2/q})}{p(p-p^{2/q})^2}$ & $-1$ & $-1$ & ${\rm Sp}(N)$ fermionic  \\ [5pt]
  \gr $\mathbb{Q}_p$ & $p \equiv 1 \mod 4$ & $p$ & $1/p$ & $-1$ & $-1$ & ${\rm Sp}(N)$ fermionic  \\ 
  \gr $\mathbb{Q}_p$ & $p \equiv 1 \mod 4$ & $\epsilon p$ & $1/p$ & $-1$ & $-1$ & ${\rm Sp}(N)$ fermionic  \\ 
  \gr $\mathbb{Q}_p$ & $p \equiv 3 \mod 4$ & $p$ & $-1/p$ & $1$ & $-1$ & ${\rm O}(N)$ fermionic  \\ 
  \gr $\mathbb{Q}_p$ & $p \equiv 3 \mod 4$ & $\epsilon p$ & $-1/p$ & $1$ & $-1$ & ${\rm O}(N)$ fermionic  \\  \hline
  $\mathbb{Q}_2$ & & $1$ & $-\frac{5\cdot 4^{1/q}-4-16^{1/q}}{2(4^{1/q}-2)^2}$ & $1$ & $1$ & ${\rm O}(N)$ bosonic  \\ [5pt]
  \gr $\mathbb{Q}_2$ & & $-1$ & $-1/4$ & $1$ & $-1$ & ${\rm O}(N)$ fermionic  \\
  \gr $\mathbb{Q}_2$ & & $2$ & $1/8$ & $-1$ & $-1$ & ${\rm Sp}(N)$ fermionic  \\
  \gr $\mathbb{Q}_2$ & & $-2$ & $-1/8$ & $1$ & $-1$ & ${\rm O}(N)$ fermionic  \\
  \gr $\mathbb{Q}_2$ & & $3$ & $-1/4$ & $1$ & $-1$ & ${\rm O}(N)$ fermionic  \\
  $\mathbb{Q}_2$ & & $-3$ & $\frac{(1+4^{1/q})(4+4^{1/q})}{2(2+4^{1/q})^2}$ & $-1$ & $-1$ & ${\rm Sp}(N)$ fermionic  \\[5pt]
  \gr $\mathbb{Q}_2$ & & $6$ & $-1/8$ & $1$ & $-1$ & ${\rm O}(N)$ fermionic  \\
  \gr $\mathbb{Q}_2$ & & $-6$ & $1/8$ & $-1$ & $-1$ & ${\rm Sp}(N)$ fermionic  \\ \hline
 \end{tabular}\end{center}\caption{Table of melonic theories, Archimedean or ultrametric. Shaded rows indicate theories for which an exact solution to the Schwinger-Dyson equation is given in subsection \ref{ExactSolution}.}\label{ExplicitResults}\end{table}

\subsection{Full solution to the Schwinger-Dyson equation for direction-dependent theories}
\label{ExactSolution}

Any $p$-adic number $x$ can be represented by a series expansion
\eqn{padicExpansion}
{
x=p^{v(x)}\sum_{m=0}^\infty c_m\, p^{m}
}
where $c_m \in \{0,1,...,p-1\}$ and $c_0 \neq 0$.

It was shown in \cite{Gubser:2017qed} for $q=4$ that the Schwinger-Dyson equation \eqref{SchwingerDysonConvolution} can be solved exactly for $p$-adic theories when the sign function is direction-dependent, that is, when the sign function $\sgn(x)$ depends not only on the norm $|x|$ but also on the first $p$-adic digit $c_0$ in the expansion \eqref{padicExpansion}. These theories are indicated by shaded rows in table \ref{ExplicitResults}, and the solutions of \cite{Gubser:2017qed} generalize straightforwardly to higher values of $q$.

For each of the theories, the solution can be written as

\eqn{}
{
G(t)=\mathcal{G}(|t|)\sgn(t) \,,
}
where $\mathcal{G}(|t|)$ is the unique real root of a $q$-th order polynomial. 

For the direction-dependent theories with odd $p$, ie. those with a sign function defined by $\tau=p$ or $\tau=\epsilon p$, $\mathcal{G}(|t|)$ is given by the real solution to the equation
\eqn{}
{
\mathcal{G}(|t|)=\sqrt{\sgn(-1)}\Gamma(\pi_{1-s,\sgn})|t|^{s-1}\left(1-\frac{\tilde{g}^2}{p}|t|^2 \,\mathcal{G}(|t|)^q \right) .
}
For $p=2$ and $\tau=-1$ or $\tau=3$, $\mathcal{G}(|t|)$ is given by the real solution to
\eqn{}
{
\mathcal{G}(|t|)=\sqrt{\sgn(-1)}\Gamma(\pi_{1-s,\sgn})|t|^{s-1}\left(1-\frac{\tilde{g}^2}{2^2}|t|^2 \,\mathcal{G}(|t|)^q \right) .
}
And $p=2$ and $\tau=\pm 2$ or $\tau = \pm 6$, $\mathcal{G}(|t|)$ is given by the real solution to
\eqn{}
{
\mathcal{G}(|t|)=\sqrt{\sgn(-1)}\Gamma(\pi_{1-s,\sgn})|t|^{s-1}\left(1-\frac{\tilde{g}^2}{2^3}|t|^2 \,\mathcal{G}(|t|)^q \right) \,.
}
In all cases, we choose $\sqrt{\sgn(-1)}$ as in \eno{sqrtSgn}.

\section{The four-point function}
\label{FOURPOINT}

\begin{figure}
 \centerline{\includegraphics[width=5in]{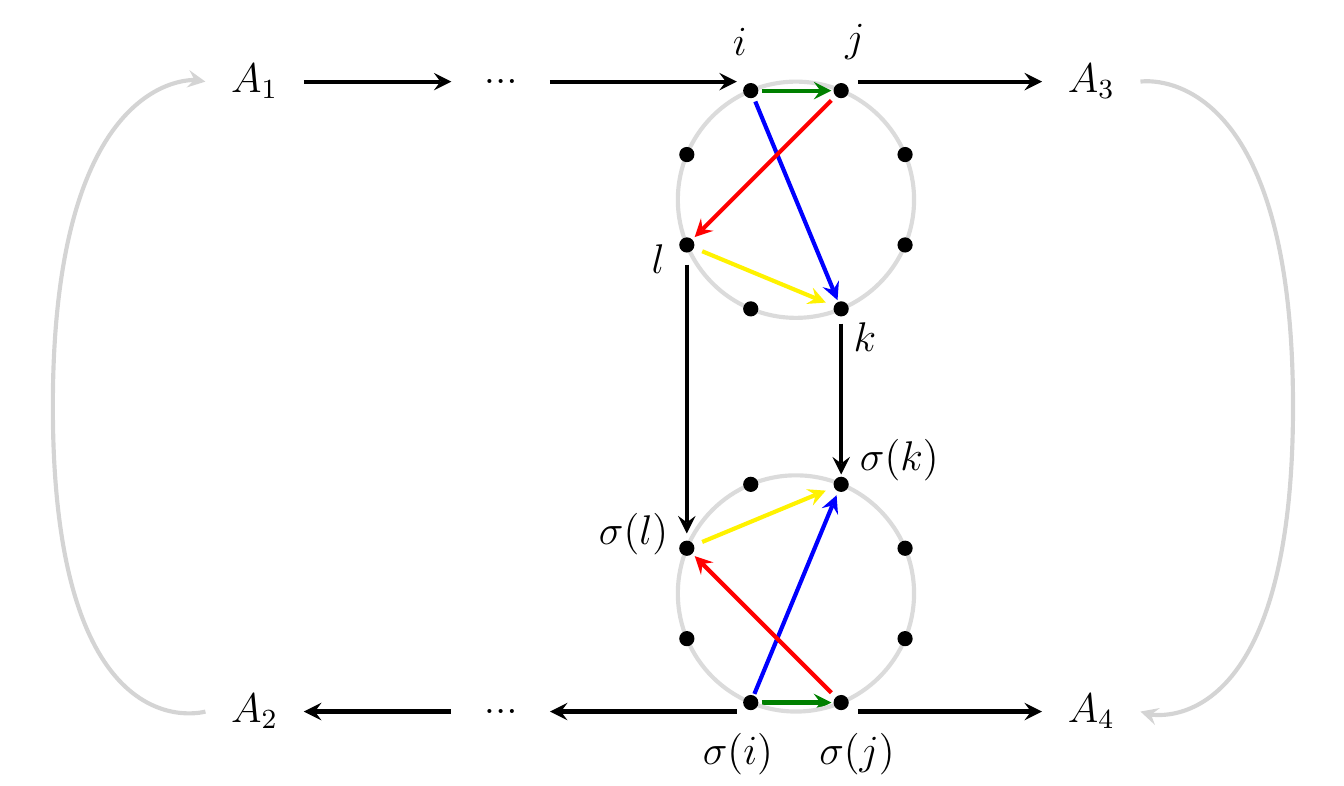}}
 \caption{Insertion of an extra set of rungs in an oriented ladder. Not all internal threads in the interactions nor all rungs have been drawn so as to not clutter up the figure.}\label{LadderRecursion}
\end{figure}

As shown in \cite{Klebanov:2016xxf}, the four-point function of the Klebanov-Tarnopolsky tensor model with rank three fermions has the same structure as the SYK model \cite{Polchinski:2016xgd}, and this result generalizes to values of $q>4$ \cite{Narayan:2017qtw}. Working to sub-leading order in the melonic limit, the four-point correlator can be decomposed as
\eqn{4pt}{
&\Omega_{A_1A_2}\Omega_{A_3A_4}\left<T(\psi^{A_1}(t_1)\psi^{A_2}(t_2)\psi^{A_3}(t_3)\psi^{A_4}(t_4))\right>
=
\cr
&N^{2(q-1)}G(t_{12})G(t_{34})+N^{q-1}\bigg(\Gamma(t_1,t_2,t_3,t_4)+\sigma_\psi\Gamma(t_1,t_2,t_4,t_3)\bigg)+\mathcal{O}(N^{q-2}) \,,
}
where $\Gamma$ stands for a sum of so-called ladder Feynman diagrams that can be expanded as $\Gamma= \sum_n \Gamma_n$ according to the number $n$ of sets of $q-2$ rungs in the ladder diagrams. Schematically, for $q=8$:
\eqn{}
{
\Gamma(t_1,t_2,t_3,t_4) =& 
\begin{matrix}\includegraphics[height=12ex]{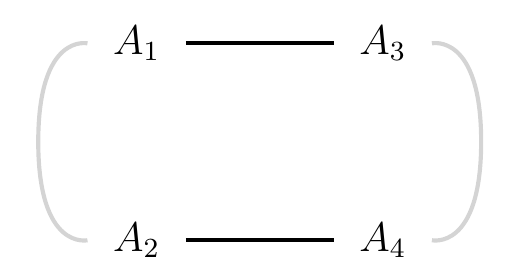}\end{matrix}+
\begin{matrix}\includegraphics[height=12ex]{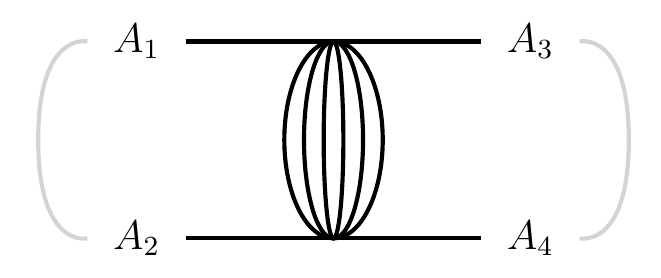}\end{matrix}+\cr
&\begin{matrix}\includegraphics[height=12ex]{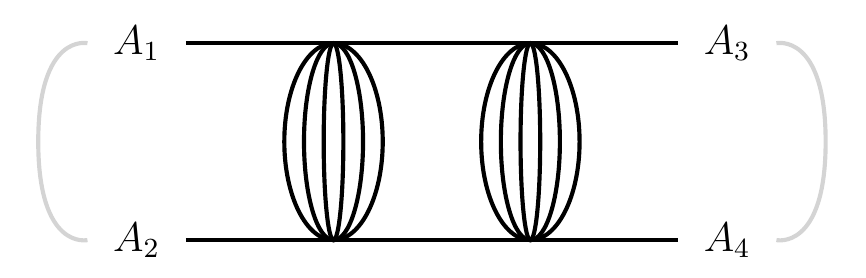}\end{matrix}+\,...
}
Here black lines represent propagators while gray lines stand for contraction through the matrix $\Omega_{AB}$. But for a more exact understanding of the four-point function, we must also consider the threads running within each interaction vertex and endow each line with an orientation. For the index contraction in \eqref{4pt}, the arrows of the first ladder diagram are oriented as
\eqn{}
{
\begin{matrix}\includegraphics[height=12ex]{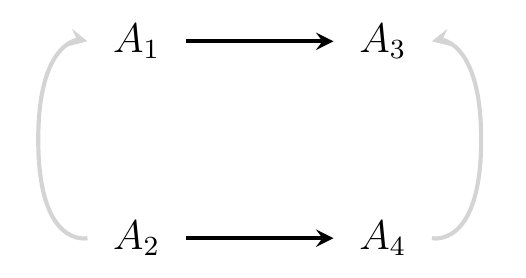}\end{matrix}.
}
By flipping two arrows, each corresponding to $q-1$ interchanges of indices in a matrix $\Omega^{ab}$ or $\Omega_{ab}$, we get an oriented diagram,
\eqn{}
{
\begin{matrix}\includegraphics[height=12ex]{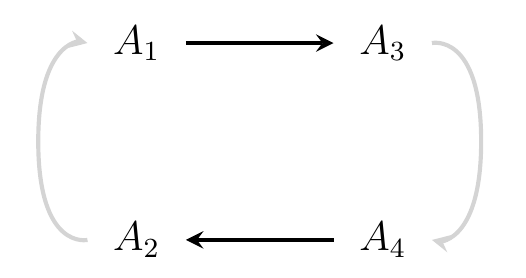}\end{matrix}.
}
Since an even number of arrow flips were performed, there is no $\sigma_\Omega$ dependence so that
\eqn{gamma0}{
\Gamma_0(t_1,t_2,t_3,t_4)=\sigma_\psi G(t_1-t_3)G(t_2-t_4)\,.
}
The absence of a factor of $\sigma_\Omega$ in \eqref{gamma0} owes directly to the choice of contraction with matrices $\Omega_{A_1A_2}\Omega_{A_3A_4}$ rather than say $\Omega_{A_1A_2}\Omega_{A_4A_3}$ in \eqref{4pt}, and so is ultimately the result of a convention. The important question to ask is whether the insertion of each new set of rungs in a ladder diagram leads to a factor of $\sigma_\Omega$. Consider therefore, as in figure~\ref{LadderRecursion}, the diagram obtained by appending an extra set of rungs to some oriented ladder diagram.

The propagator that forms part of the top rail of the ladder immediately to the left of the newly appended set of rungs may be incident to any vertex $i$ of the top interaction vertex. The propagator that forms part of the top rail immediately to the right of the appended sets of rungs may be incident to any other vertex $j\neq i $ of the topmost interaction vertex. But for a given choice of $i$ and $j$, in order to get a ladder diagram that is not suppressed in the melonic limit, the two propagators of the bottom rail immediately to the left and right of the bottom interaction vertex must be given, respectively, by $\sigma(i)$ and $\sigma(j)$ for some $\sigma\in G$. And similarly the appended rungs must connect vertex $l$ of the top interaction vertex to vertex $\sigma(l)$ of the bottom one for all $l\neq i,j$. These facts follow immediately from the above discussion of the two-point function once we note that 1) the threads of the propagators incident to vertices $i$ and $\sigma(i)$ all partake of the same index loops; and 2) from the point of view of index contraction we may look upon the propagators incident to vertices $i$ and $\sigma(i)$, the arrows running to and between $A_3$ and $A_4$, and the two appended interaction vertices and the rungs running between them as all forming part of a melonic insertion into one single propagator. Since there are $q$ choices of vertex $i$, $(q-1)$ choices of vertex $j$, and $|G|$ choices of permutation $\sigma$, the total number of melonic Feynman diagrams that contribute to the ladder diagram in question is $q(q-1)|G|$.

As in our consideration of the two-point function, we can assume that the orientation of an arrow between any two vertices $i'$ and $j'$ of the top interaction vertex is the same as the orientation of the arrow between $\sigma(i')$ and $\sigma(j')$ since automorphisms induce an even number of arrow flips. As to the $q-2$ arrows that make up the rungs, these all have the same orientation prior to flipping any arrows, as one can find by computing the four-point function via functional differentiation. As in figure~\ref{LadderRecursion} we will take these arrows to point downwards rather than upwards, but this is an arbitrary choice that does not affect the parity of arrow flips needed to make all index loops have a uniform orientation of arrows.

To determine whether appending an extra set of rungs gives rise to an overall factor of  $\sigma_\Omega$, we need to consider all the index loops involved. These fall into four types:
\begin{enumerate}
\item The index loops running from $i$ through one of the rungs of the ladder and through $\sigma(i)$ without passing through $A_3$ and $A_4$, as illustrated in blue on figure~\ref{LadderRecursion}. Because there are $q-2$ such loops and they all require the same number of arrow flips to obtain uniform orientation, no net factor of $\sigma_\Omega$ is introduced on account of these index loops.
\item The index loops running from $j$ through $A_3$ and $A_4$ and back to $j$ through one of the rungs in the ladder, illustrated in red on figure~\ref{LadderRecursion}. Again these $q-2$ loops are even in number and require the same number of arrow flips for uniform orientation, so again no net factor of $\sigma_\Omega$ is introduced.
\item The index loops that run between the two interaction vertices and consist of four threads each, two in two rungs and two within the two interaction vertices. There are $(q-2)(q-3)/2$ such loops, an example of which is illustrated in yellow on figure~\ref{LadderRecursion}. Each of these index loops consists of two threads with one orientation and two index loops with the opposite orientation, so once again no net factor of $\sigma_\Omega$ is introduced in bringing about uniform orientation.
\item Lastly, there is the index loop that runs between $i$ and $j$ and $\sigma(i)$ and $\sigma(j)$, illustrated in green on figure~\ref{LadderRecursion}. Since we are assuming that the index loops of the original ladder diagram, prior to insertion of an extra set of rungs, each had a uniform orientation of arrows, two arrows must be flipped in order to give this index loop a uniform orientation (the arrow between $\sigma(i)$ and $\sigma(j)$ and the arrow between $\sigma(j)$ and $A_4$ on figure~\ref{LadderRecursion}) and again we do not pick up any net factor of $\sigma_\Omega$.
\end{enumerate}
In summary, there are no relative sign differences in ladder diagrams between $SO(N)$ and $Sp(N)$ symmetric tensor models.

Having worked out the subtleties relating to $Sp(N)$ symmetry and the automorphism group of the interaction vertex, we are ready to write down the recursive relation describing ladder diagrams:
\eqn{}
{
\Gamma_{n+1}(t_1,t_2,t_3,t_4)= \int dt\,dt'\,K(t_1,t_2,t,t')\Gamma_n(t,t',t_3,t_4)
}
where the integration kernel is given as 
\eqn{kernel}
{
K(t_1,t_2,t,t')&=
\sigma_\psi q(q-1)|G|\left(\frac{g}{|G|}\right)^2N^{\frac{(q-1)(q-2)}{2}} G(t_1-t)G(t_2-t')G(t-t')^{q-2}
\cr
&=\sigma_\psi (q-1)\tilde{g}^2G(t_1-t)G(t_2-t')G(t-t')^{q-2}\,.
}

In the IR, we can plug in the expression for the dressed propagator derived in section \ref{sec:IRsolution} to obtain
\eqn{}{
K(t_1,t_2,t_3,t_4)
=&-\frac{\sigma_\psi\sigma_\Omega(q-1)}{\Gamma(\pi_{\frac{q-2}{q},\sgn})\Gamma(\pi_{\frac{2-q}{q},\sgn})}\pi_{-\frac{2}{q},\sgn}(t_{13})\,\pi_{-\frac{2}{q},\sgn}(t_{24})\,\pi_{\frac{2(2-q)}{q},\sgn}(t_{34})\,.
}
Following \cite{Maldacena:2016hyu} and \cite{Klebanov:2016xxf} and defining 
\eqn{}{
v(t_1,t_2)\equiv\pi_{h-\frac{2}{q},\sgn}(t_{12}),}
the integral eigenvalue-equation to solve in order to find the scaling dimensions of two-particle operators is given by
\eqn{}{v(t_1,t_2)=\frac{1}{g(h,q)}\int dt_3 \,dt_4\, K(t_1,t_2,t_3,t_4)\,v(t_3,t_4)\,.
}
Changing variables to
\eqn{}{
u \equiv t_{13}\,, \hspace{20mm} v \equiv t_{42}\,,
}
makes it manifest that the integral in the eigenvalue equation is a convolution since
\eqn{}{
&\int dt_3 \, dt_4 \, \pi_{-\frac{2}{q},\sgn}(t_{13})\,\pi_{-\frac{2}{q},\sgn}(t_{24})\,\pi_{h+\frac{2(1-q)}{q}-h,\sgn}(t_{34})=
\cr
&\sgn(-1)\int du \, dv\, \pi_{-\frac{2}{q},\sgn}(u)\,\pi_{-\frac{2}{q},\sgn}(v)\,\pi_{h+\frac{2(1-q)}{q},\sgn}(t_{12}-u-v)\,.
}
Invoking the sign constraint \eqref{sgnConstraint}, the eigenvalue equation can therefore be written as
\eqn{}{\pi_{h-\frac{2}{q},\sgn}=-\frac{(q-1)}{g(h,q)\Gamma(\pi_{\frac{q-2}{q},\sgn})\Gamma(\pi_{\frac{2-q}{q},\sgn})}\,\pi_{-\frac{2}{q},\sgn}\ast\pi_{-\frac{2}{q},\sgn}\ast\pi_{h+\frac{2(1-q)}{q},\sgn}\,.
}
Fourier-transforming this equation using \eqref{Fourier}, the multiplicative characters cancel, as do two of the gamma functions, and we find that
\eqn{}{
g(h,q)=-(q-1)\frac{\Gamma(\pi_{\frac{q-2}{q},\sgn})\Gamma(\pi_{h+\frac{2-q}{q},\sgn})}{\Gamma(\pi_{\frac{2-q}{q},\sgn})\Gamma(\pi_{h+\frac{q-2}{q},\sgn})}\,. 
}
This formula is valid for real as well as $p$-adic numbers, and for an action \eqref{SKTextended} with a kinetic term with any sign function.

On the real numbers, selecting the usual sign function in the action \eqref{SKTextended}, the above equation reproduces the fermionic result of the SYK and melonic tensor models, equations (3.73) in \cite{Maldacena:2016hyu} and (6.8) in \cite{Narayan:2017qtw}. Selecting the trivial sign on the reals, the above equation reproduces the bosonic result, equation (4.14) in \cite{Klebanov:2016xxf} with $d=1$. 

On the $p$-adic numbers, there are multiple inequivalent sign functions, each of which can be labeled by a $p$-adic number $\tau$ as explained in appendix \ref{GammaAppendix}. For the sign functions characterized by $\tau=p$ or $\tau=\epsilon p$, $g(h,q)$ reduces to $1-q$. For $\tau=\epsilon$, $g(h,q)$ is a non-constant function in $h$, but the equation $g(h,q)=1$ never has a solution for the $p$-adic theories.

\subsection{Adelic product formula for the integral eigenvalues}
\label{ADELIC}

In \cite{Ruelle:1989dg} it is demonstrated how, by invoking the functional equations of suitably chosen Dirichlet $L$-functions, one can derive an adelic product formula for the signed Gel'fand-Graev gamma functions. By selecting a fixed rational number $\tau$, one picks out a sign function for each of the number fields $\mathbb{R}$ and $\mathbb{Q}_p$. For each of these fields with associated sign function $\sgn_\tau(x)$ there is a signed character $\pi_{s,\sgn}(x)$ and an associated gamma function $\Gamma(\pi_{s,\sgn})$. Taking the product over all these gamma functions for any fixed complex number $s$, one gets the usual sign function of $\tau$:

\eqn{RuelleFormula}
{
\prod_{K=\mathbb{R},\mathbb{Q}_p \forall p}\Gamma(\pi_{s,\sgn})=\begin{cases}
1 \hspace{8mm} \text{ for }\tau >0 \cr
-1 \hspace{5mm} \text{ for }\tau <0\,.
\end{cases}
}

Note that for any choice of $\tau$ there will be many fields $K$ for which $\sgn_\tau(x)$ is the trivial sign character, ie. unity, so that $\pi_{s,\sgn}$ reduces to $\pi_s$. For example, this happens for $K=\mathbb{R}$ when $\tau$ is positive.

As an immediate consequence of \eqref{RuelleFormula}, we find that for fixed rational $\tau$, complex $h$, and even number $q$, the integral eigenvalues $g(h,q)$ satisfy the adelic product formula

\eqn{}
{
\prod_{K=\mathbb{R},\mathbb{Q}_p \forall p}
\frac{g(h,q)}{1-q}=1\,.
}
For most choices of $\tau$, this product mixes bosonic and fermionic, and $O(N)$ and $Sp(N)$ symmetric theories.

\section{Conclusions}

Given the relative uniqueness and simplicity of melonic theories with quartic and sextic vertices, it is surprising that melonic theories proliferate and diversify for larger orders $q$ of the interaction vertex.  Already at $q=8$, there are six different types of vertices, with up to $\mathbb{Z}_2^3$ symmetry.  The situation can be compared with matrix models, where if we restrict to quartic vertices, the most commonly studied interactions are $\tr \Phi^4$ and $(\tr \Phi^2)^2$.  The first of these has $\mathbb{Z}_4$ symmetry, and the second has $\mathbb{Z}_2^3$ symmetry.  If we restrict to only one of these two vertices, then we get remarkably different behavior in the large $N$ limit: $\tr \Phi^4$ leads to a planar limit, while $(\tr \Phi^2)^2$ leads to bubble diagrams.  Mixing the two gives some interesting modifications of scaling exponents of the pure $\tr \Phi^4$ theory \cite{Das:1989fq,Sugino:1994zr,Gubser:1994yb,Klebanov:1994kv}.  For melonic theories, our story so far is less featured, in that the treatment of the two-point and four-point functions proceeds almost identically for the myriad theories we can construct.  The one salient difference among the theories we consider is that the effective coupling constant that enters into the self-consistent treatment of two- and four-point functions includes the inverse half power of the order of the symmetry group of the interaction vertex.  An interesting question for future work is to see whether combining different interaction vertices could significantly alter the analysis, for example through some cancellations or modified infrared scaling. Another interesting question for future work is to examine the extent to which the operator counting of \cite{Bulycheva:2017ilt} generalizes to higher melonic theories.

Our construction of $q$-fold interaction vertices is, by necessity, somewhat detailed, amounting to a coloring of the complete graph of $q$ vertices with $q-1$ colors so that all edges meeting at a vertex have different colors.  The combinatorial problem of counting all such colorings is formidable, and it is related to the problem of one-factorizations of the complete graph, for which results are available only for modest values of $q$.  We have however demonstrated the existence of colorings with all possible symmetries groups, namely $\mathbb{Z}_2^n$ with $n$ bounded above by $v$ if $q = 2^v$ and by $v-1$ if $q = u 2^v$ with odd $u$ greater than $1$.  While our construction of interaction vertices is rooted in the natural generalizations of the Klebanov-Tarnopolsky model \cite{Klebanov:2016xxf}, we cannot claim to have exhausted all possible generalizations with melonic limits.  Here is a useful way to frame the question: If we restrict the field content to a single real field $\psi^{a_0a_1\dots a_{q-2}}$ with $q-1$ $N$-valued indices all of which must be contracted amongst each other, is the most general interaction vertex with a melonic limit a linear combination of the interaction vertices that can be described in terms of colorings of a complete graph of $q$ vertices?  We hope to see progress on this and related issues in future work.

\subsection*{Acknowledgments}

We thank Igor Klebanov for useful discussions and suggestions. This work was supported in part by the Department of Energy under Grant No.~DE-FG02-91ER40671, and by the Simons Foundation, Grant 511167 (SSG).  The work of C.J.\ was also supported in part by the US NSF under Grant No.~PHY-1620059.

\appendix

\section{Gel'fand-Graev gamma functions}
\label{GammaAppendix}

In this appendix we review the Gel'fand-Graev gamma functions associated with certain additive and multiplicative characters defined over a locally compact field $K$ that can be either $\mathbb{R}$, $\mathbb{C}$, or $\mathbb{Q}_p$ and with an associated norm $|\cdot|$ that can be either the absolute value or the $p$-adic norm. 

For a given additive character $\chi\colon K \rightarrow \mathbb{C}$, the Gel'fand-Graev gamma function associated with a multiplicative character $\pi\colon K \rightarrow \mathbb{C}$ is defined by
\eqn{gammaDef}
{\Gamma(\pi)=\int_K \frac{dx}{|x|}\,\chi(x)\,\pi(x)\,.}

The additive character $\chi\colon K \rightarrow \mathbb{C}$  we are working with is
\eqn{}
{
\chi(x) \equiv \begin{cases} e^{2\pi i x} \hspace{7.3mm} \text{for }K=\mathbb{R},\mathbb{C} \cr e^{2\pi i \{x\}} \hspace{5mm} \text{for }K=\mathbb{Q}_p \end{cases}
}
where $\{x\}$ denotes the fractional part of $x$.

The multiplicative characters that are relevant to this paper are the ones given in equation \eqref{multiplicativecharacters},
\eqn{multiplicativecharacters2}
{
\pi_s(t) \equiv |t|^s\,, \hspace{10mm} \pi_{s,\sgn}(t) \equiv |t|^s\sgn(t)\,.
}

Defining the Fourier transform $\mathcal{F}[f]\colon K\rightarrow \mathbb{C}$ of a function $f\colon K\rightarrow \mathbb{C}$ by
\eqn{}
{
\mathcal{F}[f](\omega)=\int_K dx\,\chi(x\omega)f(x)\,,
}
it is straightforward to show using the definition \eqref{gammaDef} that the Fourier transforms of the multiplicative characters \eqref{multiplicativecharacters2} are given by
\eqn{Fourier2}
{
\mathcal{F}[\pi_s](\omega)=\Gamma(\pi_{s+1})\,\pi_{-s-1}(\omega)\,, \hspace{10mm}
\mathcal{F}[\pi_{s,\sgn}](\omega)=\Gamma(\pi_{s+1,\sgn})\,\pi_{-s-1,\sgn}(\omega)\,.
}

In order to write down explicit expressions for the Gel'fand-Graev gamma functions associated with $\mathbb{R}$ and $\mathbb{Q}_p$ and the multiplicative characters \eqref{multiplicativecharacters2}, it is expedient to introduce the local zeta functions $\zeta_\infty,\zeta_p\colon \mathbb{C} \rightarrow \mathbb{C}$:

\eqn{}
{
\zeta_\infty(s)=\pi^{-\frac{s}{2}}\,\Gamma_\text{E}\left(\frac{s}{2}\right), \hspace{15mm} \zeta_p(s)=\frac{1}{1-p^{-s}}\,,
}
where $\Gamma_\text{E}(s)$ is the familiar Euler gamma function. 

For $K=\mathbb{R}$, the Gel'fand-Graev gamma functions are given by

\eqn{}
{
\Gamma(\pi_s)=\frac{\zeta_\infty(s)}{\zeta_\infty(1-s)}\,, \hspace{15mm} \Gamma(\pi_{s,\sgn})=i\frac{\zeta_\infty(1+s)}{\zeta_\infty(2-s)}\,.
}

For $K=\mathbb{C}$ there are no sign functions, and the Gel'fand-Graev gamma function is given by

\eqn{}
{
\Gamma(\pi_s)=(2\pi)^{-2s}\left(\Gamma_E(s)\right)^2\sin(\pi s)\,.
}

For $K=\mathbb{Q}_p$, there are multiple inequivalent sign functions each associated with a quadratic extension $\mathbb{Q}_p(\sqrt{\tau})$ formed by the field extension of $\mathbb{Q}_p$ of with respect to $\sqrt{\tau}$, where $\tau$ is some $p$-adic number that is not a square of a $p$-adic number. For any $z \in \mathbb{Q}_p(\sqrt{\tau})$, there exists unique $x,y \in \mathbb{Q}_p$ such that 
\eqn{}
{
z = x+\sqrt{\tau}y\,.
}
For any $z \in  \mathbb{Q}_p(\sqrt{\tau})$ it is therefore possible to define a conjugate element $z^\ast \in  \mathbb{Q}_p(\sqrt{\tau})$ by
\eqn{}
{
z^\ast = x-\sqrt{\tau}y\,,
}
and for any quadratic extension $\mathbb{Q}_p(\sqrt{\tau})$ one can define a sign function $\sgn_\tau\colon \mathbb{Q}_p^\times\rightarrow\{1,-1\}$ by
\eqn{}
{\sgn_\tau(x)=\begin{cases} 1 \hspace{7mm}\text{ if there exists }  z\in\mathbb{Q}_p(\sqrt{\tau}) \text{ such that } x=z^\ast z
\cr -1 \hspace{5mm} \text{otherwise}\,.  \end{cases}}
An obvious point is that if $\tilde\tau = \tau a^2$, where $\tau$, $\tilde\tau$, and $a$ are all non-vanishing $p$-adic numbers, then $\tau$ and $\tilde\tau$ define the same sign character.

For each sign function, labeled by a $p$-adic number $\tau$, we can define a multiplicative character
\eqn{}
{
\pi_s^{(\tau)}(x)\equiv |x|^s\sgn_\tau(x) \,.
}

This definition extends to $x\in \mathbb{R}$ by letting $\sgn_\tau(x)$ denote the usual sign function for $\tau<0$ and the trivial sign function $\sgn_\tau(x)=1$ for $\tau>0$.

For $p=2$, there are seven distinct non-trivial sign functions corresponding to $\tau = -1$, $\pm 2$, $\pm 3$, and $\pm 6$, and the gamma functions evaluate to
\eqn{}{
&\Gamma_2(\pi_s)=\frac{\zeta_2(s)}{\zeta_2(1-s)}\,,\cr
&\Gamma_2(\pi_s^{(-3)})=\frac{\zeta_2(1-s)\zeta_2(2s)}{\zeta_2(2-2s)\zeta_2(s)}\,,\cr
&\Gamma_2(\pi_s^{(-1)})=\Gamma_2(\pi_s^{(3)})=i\frac{4^s}{2}\,,\cr
&\Gamma_2(\pi_s^{(-2)})=-\Gamma_2(\pi_s^{(6)})=i\frac{8^s}{\sqrt{8}}\,,\cr
&\Gamma_2(\pi_s^{(2)})=-\Gamma_2(\pi_s^{(-6)})=\frac{8^s}{\sqrt{8}}\,.
}

For $p>2$ there are three distinct non-trivial sign functions, which can be labeled by $\tau$ equal to $p$, $\epsilon$, and $\epsilon p$, where $\epsilon$ is an integer that is not a square modulo $p$. The Gel'fand-Graev gamma functions are given by
\eqn{}{
&\Gamma_{p}(\pi_s)=\frac{\zeta_p(s)}{\zeta_p(1-s)}\,,
\hspace{20mm}\Gamma_{p}(\pi_s^{(\epsilon)})=\frac{\zeta_p(1-s)\zeta_p(2s)}{\zeta_p(2-2s)\zeta_p(s)}\,,
\cr
\cr
&\Gamma_p(\pi_s^{(p)})=\begin{cases}\displaystyle\frac{p^s}{\sqrt{p}}\,,\cr\displaystyle 
-i\,\frac{p^s}{\sqrt{p}}\,,
\end{cases}
\hspace{5mm}
\Gamma_p(\pi_s^{(\epsilon p)})=\begin{cases}\displaystyle -\frac{p^s}{\sqrt{p}}  \hspace{9mm} \text{for }p\equiv 1 \mod 4\,,\cr
\displaystyle i\, \frac{p^s}{\sqrt{p}},  \hspace{10mm} \text{for }p\equiv 3 \mod 4\,.\end{cases}
}

\section{Finding the number of isomorphism classes of one-factorizations for \texorpdfstring{$q=12$}{q=12} using the orderly algorithm}
\label{AlgorithmAppendix}

In this appendix we discuss in detail how to find the numbers of melonic interactions with trivial and non-trivial automorphism groups for $q=12$.

We have seen that melonic interaction vertices can be identified with isomorphism classes of one-factorizations of complete graphs. For $q=12$ the total number of isomorphism classes was found in \cite{Dinitz:1994}. To classify melonic theories it is desirable to also know how many of these isomorphism classes correspond to melonic interactions with a non-trivial automorphism group. A one-factorization corresponds to such an interaction if there exists a vertex permutation $\pi \in S_q^{\text{ vertices}}$ that leaves all one-factors invariant. We will now show, for $q=12$, how to employ the orderly algorithm of \cite{Seah} and \cite{Dinitz:1994} to find the total number of isomorphism classes of one-factorizations as well as the number of isomorphism classes corresponding to melonic interactions with non-trivial automorphisms.

A one-factor $F_i$ can be represented by a set of six pairs
\eqn{OneFactor}
{
F_i=\big\{\{a_0,a_1\},\{a_2,a_3\},\{a_4,a_5\},\{a_6,a_7\},\{a_8,a_{9}\},\{a_{10},a_{11}\}\big\}
}
where all the $a_i$ are distinct and belong to the set $\{0,1,...,11\}$. 

By forming a set of $I$ pair-wise disjoint one-factors, we get a partial one-factorization $H_I$,
\eqn{OneFactorization}
{
H_I=\{F_0,F_1,...,F_{I-1}\}\,
}
where $1\leq I < 12$. A (full) one-factorization is a partial one-factorization with $I=11$. In the notation of section \ref{ONEFACTOR}, $H_{11}=\overline{\mathcal{F}}$.

A key point in implementing the orderly algorithm consists in imposing a lexicographic order amongst partial one-factorizations. One first imposes an order amongst the one-factors as follows:
\begin{itemize}
\item Associate to each one-factor $F_i$ a unique ordered $12$-tuple $\tilde{F}_i$ by writing it in the form \eqref{OneFactor} with $a_{2i}<a_{2i+1}$ and $a_{2i}<a_{2i+2}$ for all $i$. Then take $\tilde{F}$ to be given by
\eqn{}
{(a_0,a_1,a_2,a_3,a_4,a_5,a_6,a_7,a_8,a_9,a_{10},a_{11})\,.
}
\item Given two distinct one-factors $F_i$ and $F'_i$, we impose an order such that $F_i<F'_i$ if the left-most digit $a_i$ by which $\tilde{F}_i$ and $\tilde{F}'_i$ differ is smaller for $\tilde{F}_i$.  
\end{itemize}
Having ordered the one-factors, we can now order the partial one-factorizations:
\begin{itemize}
\item Associate to each partial one-factorization $H_I$ a unique ordered $I$-tuple $\tilde{H}_I$ by writing it in the form \eqref{OneFactorization} with $F_{i}<F_{i+1}$ for all $i$. Then take $\tilde{H}_I$ to be given by
\eqn{}
{(\tilde{F}_0,\tilde{F}_1,...,\tilde{F}_{I-1})\,.
}
\item Given two distinct partial one-factorizations $H_I$ and $H'_I$, we impose an order such that $H_I<H'_I$ if the left-most digit $a_i$ by which $\tilde{H}_I$ and $\tilde{H}'_I$ differ is smaller for $\tilde{H}_I$.  
\end{itemize}
Now, given a partial one-factorization, we can act on it with a permutation $\pi \in S_{12}$ by permuting the digits of all the one-factors:
\eqn{}
{
a_i \rightarrow \pi(a_i)\,.
}
With these preliminaries in place, we are ready to describe the orderly algorithm. The algorithm works by constructing non-isomorphic partial one-factorizations $H_I$ and then using these to construct larger non-isomorphic partial one-factorizations $H_{I+1}$. The way one ensures that all constructed partial one-factorizations $H_I$ are non-isomorphic is by always selecting from any isomorphism class only the one with the lowest lexicographic ordering. The steps of the algorithm can be described as follows:
 \begin{enumerate}
  \item Consider all partial one-factorizations $H_1$ containing exactly one-factor and discard the ones for which there exists a permutation $\pi \in S_{12}$ such that $\pi H_1 < H_1$. Denote the set of remaining partial one-factorizations $H_1$ by $\mathcal{H}_1$.\label{FirstStep}
  \item Consider all partial one-factorizations $H_2$ obtained by adjoining to each member $H_1$ of $\mathcal{H}_1$ a one-factor disjoint from the one-factor in $H_1$. Discard the ones for which there exists a permutation $\pi \in S_{12}$ such that $\pi H_2 < H_2$. Denote the set of remaining partial one-factorizations $H_2$ by $\mathcal{H}_2$.\label{SecondStep}
  \item Continue on as in step~\ref{SecondStep} to obtain sets $\mathcal{H}_n$ of ordered partial one-factorizations for $n=3,4,\dots,10$.\label{ContinuingStep}
  \item Consider all one-factorizations $H_{11}$ obtained by adjoining to each member $H_{10}$ of $\mathcal{H}_{10}$ a one-factor disjoint from all the one-factors in $H_{10}$. Discard the ones for which there exists a permutation $\pi \in S_{12}$ such that $\pi H_{11} < H_{11}$. Denote the set of remaining one-factorizations $H_{11}$ by $\mathcal{H}_{11}$. This set will contain exactly one one-factorizations for each isomorphism class of one-factorizations, namely the one with the lowest lexicographic ordering.\label{LastStep}
 \end{enumerate}

The efficiency of the above algorithm can be improved in several ways. In considering in step~\ref{FirstStep} the partial one-factorizations $H_1$ containing one one-factor, one can restrict consideration to one-factors containing the pair $\{0,1\}$. For the one-factors adjoined to the members of $\mathcal{H}_1$ in step~\ref{SecondStep}, one can restrict consideration to one-factors containing the pair $\{0,2\}$. Similarly one can consider restricted sets of one-factors in implementing the rest of step~\ref{ContinuingStep} as well as step~\ref{LastStep}, where one needs only adjoin one-factors containing $\{0,11\}$ to the members of $\mathcal{H}_{10}$.

The algorithm as described above will find all isomorphism classes of one-factorizations as was done in \cite{Dinitz:1994}. To find only the isomorphism classes corresponding to melonic interactions with non-trivial automorphism groups, the algorithm must be modified. As we have seen in section \ref{SYMMETRY}, for $q=12$ the only possible non-trivial automorphism group is $\mathbb{Z}_2$. So for any one-factorization there can be at most a single non-trivial permutation $\pi \in S_{12}$ that leaves all one-factors invariant, and we have also seen in section \ref{SYMMETRY} that such a permutation must consist of six two-cycles. Let $\tilde{S}_{12}$ denote the subset of permutations in $S_{12}$ that consist of six two-cycles. In total there are 11!! such permutations.

For any one-factorization $H_{11}$ whose one-factors are invariant under a permutation $\pi \in \tilde{S}_{12}$, one can act on it with some permutation $\sigma\in S_{12}$ to generate a one-factorization $\sigma H_{11}$ whose one-factors are invariant wrt. any other permutation $\pi' \in \tilde{S}_{12}$. The isomorphism classes of one-factorizations corresponding to a melonic vertex with $G=\mathbb{Z}_2$ each contains a number of elements divisible by 11!! as each of them contains the same number of one-factorizations for each member of $\tilde{S}_{12}$. For the purpose of finding the number of such isomorphism classes by constructing exactly one example from each class, we can therefore restrict attention to one-factors invariant under one specific permutation $\tilde{\pi} \in \tilde{S}_{12}$, say
\eqn{SpecificPermutation}
{
\tilde{\pi}=(0,1)(2,3)(4,5)(6,7)(8,9)(10,11)\,.
}
The modifications by which we need to adjust steps~\ref{FirstStep} to~\ref{LastStep} above, then, consist of:
\begin{itemize}
\item Throughout the implementation of the algorithm, one should only consider one-factors that are invariant under the permutation \eqref{SpecificPermutation}. In total there are 331 such one-factors.
\item Throughout steps 1 to 4, when checking for each partial one-factorization $H_I$ whether there exists a permutation $\pi$ such that $\pi H_I < H_I$, one should only consider permutations $\pi$ that respect the symmetry under \eqref{SpecificPermutation}, ie.~permutations that only mix the 331 one-factors invariant under $\tilde{\pi}$ amongst themselves, i.e.~permutations that commute with $\tilde{\pi}$. This subgroup of $S_{12}$, the centralizer $C_{S_{12}}(\tilde{\pi})$, contains 
46,080 $=\frac{12!}{11!!}=6!\cdot 2^6$ permutations.
\end{itemize}
These two restrictions make the orderly algorithm run much faster. One finds that the total number of isomorphism classes corresponding to $G=\mathbb{Z}_2$ melonic interaction vertices is 4851. Having generated one representative of each of these classes, one can act on the representatives with permutations in $\tilde{S}_{12}$ and delete duplicates to find that there are 97,032,192 one-factorizations consisting of one-factors that are each invariant under \eqref{SpecificPermutation}. Multiplying this number with 11!! one finds the total number of one-factorizations all of whose one-factors are invariant with respect to some non-trivial permutation, namely $\text{1,008,649,635,840}$.

An alternative way to obtain this number of one-factorizations is to use the orbit-stabilizer theorem in the form \eqref{OrbSum}. Acting on the 4851 non-isomorphic one-factorizations with the centralizer $C_{S_{12}}(\tilde{\pi})$, one can find the order of the automorphism group $\mathrm{Aut}(\overline{\mathcal{F}})$ for each of these one-factorizations. One can then carry out the sum

\eqn{OrbSumAgain}
{
\sum_{i=1}^{4851}\frac{|S_{12}|}{|\mathrm{Aut}(\overline{\mathcal{F}}_i)|}\, ,
}
and recover the number $\text{1,008,649,635,840}$. The terms in the sum \eqref{OrbSumAgain} assume but ten different values, and the sum can be evaluated straightforwardly using the data in table~\ref{AutTable}, which lists the number of non-isomorphic one-factorizations corresponding to interaction vertices with $\mathbb{Z}_2$ symmetry for any order of $\mathrm{Aut}(\overline{\mathcal{F}})$.

\begin{table}
\centering
\begin{tabular}{|c|c|c|c|c|c|c|c|c|c|c|}\hline
$|\mathrm{Aut}(\overline{\mathcal{F}})|$ & 2 & 4 & 6 & 8 & 10 & 12 & 16 & 24 & 48 & 240 \\\hline
$n$ & 3697 & 944 & 13 & 104 & 1 & 38 & 36 & 10 & 6 & 2
\\\hline
\end{tabular}
\caption{The number of non-isomorphic one-factorizations $n$ with $|\mathrm{Aut}(\mathcal{F})|=2$ for each automorphism group order.}
\label{AutTable}
\end{table}

\section{Isomorphism classes of ordered one-factorizations for \texorpdfstring{$q=12$}{q=12}}
\label{IsomorphismOrdered}

\begin{figure}[h!]
  \centerline{\includegraphics[width=6in]{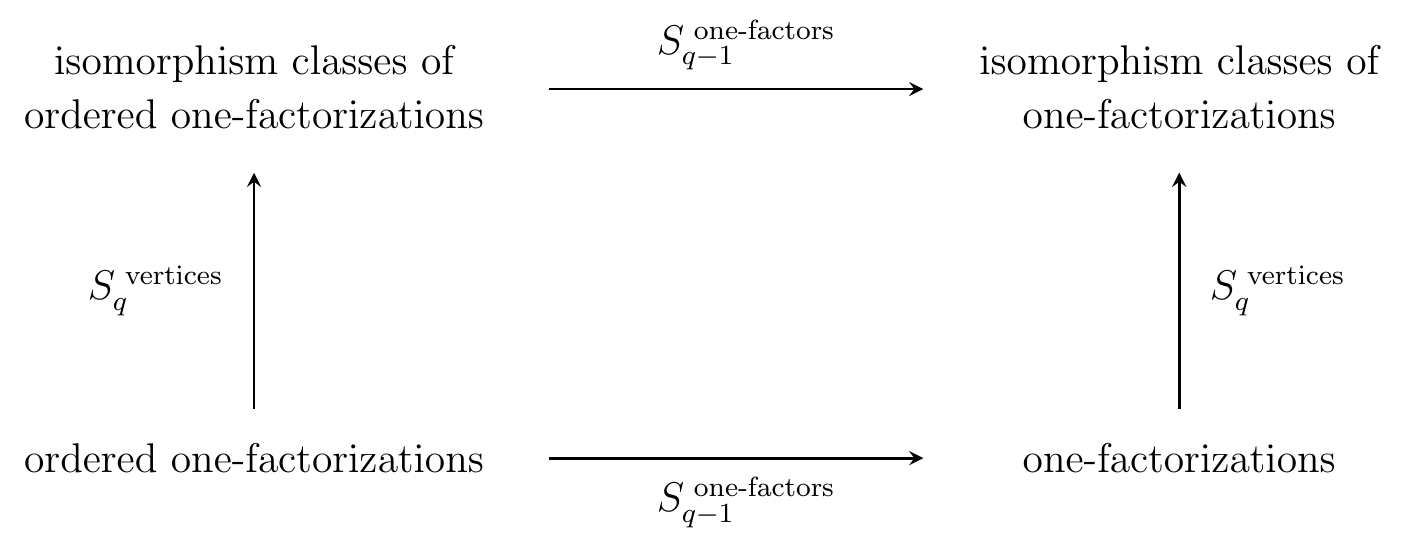}}
  \caption{The relationship between ordered one-factorizations, one-factorizations, and their isomorphism classes under vertex permutations. An arrow labeled by a group indicates that the set at the head end can be identified with the orbits of the set at the tail end under the action of the group.}\label{equivalenceBox}
 \end{figure}

For a complete graph of order $q$, we can consider the set of all ordered one-factorizations. The orbits of this set under permutations $\tau \in S_{q-1}$ of the one-factors are equal to the one-factorizations of the graph. The orbits of the one-factorizations under permutations $\pi \in S_{q}$ of the vertices are equal to the isomorphism classes that we identified with melonic interaction vertices and tabulated in figure \ref{OFTable}. But another set one can consider is the orbits of ordered one-factorizations under permutations $\pi \in S_{q}$ of the vertices. The relations between these sets and orbits are depicted diagrammatically in figure \ref{equivalenceBox}.

We argued in section \ref{SYMMETRY} that while an ordered one-factorization of a graph can immediately be translated into a melonic interaction vertex, permutations of the vertices of the graph correspond to commuting fields past each other, and permutations of the one-factors correspond to field redefinitions, neither of which operations yield physically distinct theories, so that in counting the number of theories via ordered one-factorizations one should mod out by $S_q^{\text{ vertices}}$ and $S_{q-1}^{\text{ one-factors}}$. But this reasoning only applies to lagrangians with a single interaction term. For theories with multiple interaction terms, one can still commute the fields around in each interaction term, and so one should still mod out by $S_q^{\text{ vertices}}$. But field-redefinitions can only be applied globally, and so one should no longer count the number of interaction terms by modding out by $S_{q-1}^{\text{ one-factors}}$.  A crude upper bound on the number of distinct theories with $n$ interaction vertices of order $q$ is the $n$-th power of the number of isomorphism classes of ordered one-factorizations.  The number of isomorphism classes of ordered one-factorizations, call it $C$, is therefore not without interest in the context of melonic models. In the remainder of this appendix we will derive this number in the case of $q=12$ and for interaction vertices with and without trivial automorphism groups.

Letting $X$ denote a set acted on by a finite group $G$, $X^g$ denote the set of elements in $X$ that are left invariant by $g\in G$ , and $X/G$ denote the set of orbits of $X$ under $G$, Burnside's lemma tells us that
\eqn{Burnside}
{
|X/G|=\frac{1}{|G|} \sum_{g\in G} |X^g|\,.
}
By taking $X$ to be the set of ordered one-factorizations and $G$ to be $S_{12}$ acting on vertices, we can use Burnside's lemma to find $C$. If in the sum over $g\in G$ in \eqref{Burnside} we take $g$ to be the identity element, then $|X^g|$ is equal to $|X|$ itself, that is the total number of ordered one-factorizations, which in turn is equal to 11!$M$, where $M$ is the total number of one-factorizations. From \cite{Dinitz:1994} we know that
\eqn{}
{
M=\text{252,282,619,805,368,320}\,.
}
If in the sum over $g\in G$ in \eqref{Burnside} we consider the sum over the permutations $g\in \tilde{S}_{12}$, with $\tilde{S}_{12}$ defined in appendix \ref{AlgorithmAppendix}, then this partial sum will count each ordered one-factorization whose one-factors are all invariant under some permutation $g\in \tilde{S}_{12}$ exactly once. This number is equal to 11!$M_2$, where $M_2$ is the total number of one-factorizations whose one-factors are all invariant under some permutation $g\in \tilde{S}_{12}$. We found in appendix \ref{AlgorithmAppendix} that this number is equal to
\eqn{}
{
M_2=\text{1,008,649,635,840}\,.
}
If in the sum over $g\in G$ in \eqref{Burnside}, $g$ is not the identity element nor belongs to $\tilde{S}_{12}$, then $|X^g|=0$. 

In conclusion, Burnside's lemma tells us that
\eqn{}
{
C=\frac{1}{12!}(11!\,M+11!\,M_2)=\text{21,023,635,704,583,680}\,.
}
This, the number of isomorphism classes of ordered one-factorizations, can be split into $C_1$, the number of classes corresponding to interaction vertices with trivial automorphism group, and $C_2$, the number of classes corresponding to interaction vertices with automorphism group $\mathbb{Z}_2$,
\eqn{}{C=C_1+C_2\,.}
To find out how $C$ splits into $C_1$ and $C_2$ we can use the orbit-stabilizer theorem in the form of \eqref{OrbSum} and \eqref{OrbSumAgain},
\eqn{}
{
|X|=\sum_{\text{orbits }o}\frac{|G|}{|\text{stab}(o)|}\,.
}
The orbits of ordered one-factorizations under $S_{12}$ acting on vertices fall into two sets. There are $C_1$ orbits $o$ for which $|\text{stab}(o)|=1$ and $C_2$ orbits $o$ for which $|\text{stab}(o)|=2$. Hence,
\eqn{}
{
11!M =12!\left(\frac{C-C_2}{1}+\frac{C_2}{2}\right)\,.
}
Solving for $C_2$, one finds that
\eqn{}
{
C_2=\text{168,108,272,640}\,.
}
The numbers of ordered and un-ordered one-factorizations and the numbers of isomorphism classes of these under vertex permutations for $q=12$ are summarized in the table in table~\ref{OFTable2}.\bigskip

\begin{table}[h]
\centering
\begin{tabular}{|c|c|c|}\hline
\shortstack{\\automorphism group of \\ the interaction vertex } & \shortstack{trivial\\ \vspace{1mm}} & \shortstack{$\mathbb{Z}_2$\\ \hspace{1mm}}  \\ \hline
\shortstack{\\ordered one-factorizations} &  $11!\,\cdot \text{252,281,611,155,732,480}$ &  $11!\,\cdot \text{1,008,649,635,840}$  \\\hline
\shortstack{\\one-factorizations}     &  252,281,611,155,732,480 &  1,008,649,635,840  \\\hline
\shortstack{\\isomorphism classes of \\ ordered one-factorizations} & 
21,023,467,596,311,040 &  168,108,272,640 \\\hline
\shortstack{\\isomorphism classes of \\ one-factorizations} & 526,910,769 &  4851 \\\hline
\end{tabular}
\caption{The number of equivalence classes of ordered one-factorizations under permutations of one-factors and vertices for $q=12$.}
\label{OFTable2}
\end{table}

\clearpage

\bibliographystyle{utphys}
\bibliography{higher}

\providecommand{\href}[2]{#2}\begingroup\raggedright\begin{thebibliography}{10}

\bibitem{Bonzom:2011zz}
V.~Bonzom, R.~Gurau, A.~Riello, and V.~Rivasseau, ``{Critical behavior of
  colored tensor models in the large N limit},''
  \href{http://dx.doi.org/10.1016/j.nuclphysb.2011.07.022}{{\em Nucl. Phys.}
  {\bfseries B853} (2011) 174--195},
\href{http://arxiv.org/abs/1105.3122}{{\ttfamily arXiv:1105.3122 [hep-th]}}.

\bibitem{Gurau:2011xp}
R.~Gurau and J.~P. Ryan, ``{Colored Tensor Models - a review},''
  \href{http://dx.doi.org/10.3842/SIGMA.2012.020}{{\em SIGMA} {\bfseries 8}
  (2012) 020},
\href{http://arxiv.org/abs/1109.4812}{{\ttfamily arXiv:1109.4812 [hep-th]}}.

\bibitem{Carrozza:2015adg}
S.~Carrozza and A.~Tanasa, ``{$O(N)$ Random Tensor Models},''
  \href{http://dx.doi.org/10.1007/s11005-016-0879-x}{{\em Lett. Math. Phys.}
  {\bfseries 106} no.~11, (2016) 1531--1559},
\href{http://arxiv.org/abs/1512.06718}{{\ttfamily arXiv:1512.06718 [math-ph]}}.

\bibitem{Witten:2016iux}
E.~Witten, ``{An SYK-Like Model Without Disorder},''
\href{http://arxiv.org/abs/1610.09758}{{\ttfamily arXiv:1610.09758 [hep-th]}}.

\bibitem{Klebanov:2016xxf}
I.~R. Klebanov and G.~Tarnopolsky, ``{Uncolored random tensors, melon diagrams,
  and the Sachdev-Ye-Kitaev models},''
  \href{http://dx.doi.org/10.1103/PhysRevD.95.046004}{{\em Phys. Rev.}
  {\bfseries D95} no.~4, (2017) 046004},
\href{http://arxiv.org/abs/1611.08915}{{\ttfamily arXiv:1611.08915 [hep-th]}}.

\bibitem{Sachdev:1992fk}
S.~Sachdev and J.~Ye, ``{Gapless spin fluid ground state in a random, quantum
  Heisenberg magnet},''
  \href{http://dx.doi.org/10.1103/PhysRevLett.70.3339}{{\em Phys. Rev. Lett.}
  {\bfseries 70} (1993) 3339},
\href{http://arxiv.org/abs/cond-mat/9212030}{{\ttfamily arXiv:cond-mat/9212030
  [cond-mat]}}.

\bibitem{Kitaev:2015zz}
A.~Kitaev.
\newblock ``A simple model of quantum holography.'' Talks at KITP, April 7,
  2015 and May 27, 2015.

\bibitem{Narayan:2017qtw}
P.~Narayan and J.~Yoon, ``{SYK-like Tensor Models on the Lattice},''
  \href{http://dx.doi.org/10.1007/JHEP08(2017)083}{{\em JHEP} {\bfseries 08}
  (2017) 083},
\href{http://arxiv.org/abs/1705.01554}{{\ttfamily arXiv:1705.01554 [hep-th]}}.

\bibitem{Choudhury:2017tax}
S.~Choudhury, A.~Dey, I.~Halder, L.~Janagal, S.~Minwalla, and R.~Poojary,
  ``{Notes on Melonic $O(N)^{q-1}$ Tensor Models},''
\href{http://arxiv.org/abs/1707.09352}{{\ttfamily arXiv:1707.09352 [hep-th]}}.

\bibitem{Ferrari:2017jgw}
F.~Ferrari, V.~Rivasseau, and G.~Valette, ``{A New Large N Expansion for
  General Matrix-Tensor Models},''
\href{http://arxiv.org/abs/1709.07366}{{\ttfamily arXiv:1709.07366 [hep-th]}}.

\bibitem{Tarnopolsky:2018env}
G.~Tarnopolsky, ``{On large $q$ expansion in the Sachdev-Ye-Kitaev model},''
\href{http://arxiv.org/abs/1801.06871}{{\ttfamily arXiv:1801.06871 [hep-th]}}.

\bibitem{prethesis}
C.~Jepsen, ``{Adelic Tensor Models},'' {\em Princeton University Pre-thesis}
  (2017) .

\bibitem{Dickson:1906}
L.~E. Dickson and F.~H. Safford, ``{Solution to problem 8 (group theory)},''
  {\em Amer. Math. Monthly} {\bfseries 13} (1906) 150--151.

\bibitem{Gelling:1973}
E.~N. Gelling, ``{On one-factorizations of a complete graph and the
  relationship to round-robin schedules},'' 1973.
\newblock Ph.D. Thesis, Univ. Victoria, Canada.

\bibitem{Gelling:1974}
E.~N. Gelling and R.~E. Odeh, ``{On 1-factorizations of a complete graph and
  the relationship to round-robin schedules},'' {\em Congressus Numerantium}
  {\bfseries 9} (1974) 213--221.

\bibitem{Dinitz:1987}
J.~H. Dinitz and D.~R. Stinson, ``{A hill-climbing algorithm for the
  construction of one-factorizations and Room squares},'' {\em SIAM Journal on
  Algebraic Discrete Methods} {\bfseries 8} no.~3, (1987) 430--438.

\bibitem{Seah}
E.~Seah and D.~R. Stinson, ``{On the enumeration of one-factorizations of
  complete graphs containing prescribed automorphism groups},'' {\em
  Mathematics of computation} {\bfseries 50} no.~182, (1988) 607--618.

\bibitem{Dinitz:1994}
J.~H. Dinitz, D.~K. Garnick, and B.~D. McKay, ``{There are 526,915,620
  nonisomorphic one-factorizations of K12},'' {\em Journal of Combinatorial
  Designs} {\bfseries 2} no.~4, (1994) 273--285.

\bibitem{Kaski:2009}
P.~Kaski and P.~R. \"Osterg{\aa}rd, ``{There are 1,132,835,421,602,062,347
  nonisomorphic one-factorizations of K14},'' {\em Journal of Combinatorial
  Designs} {\bfseries 17} no.~2, (2009) 147--159.

\bibitem{Gubser:2017qed}
S.~S. Gubser, M.~Heydeman, C.~Jepsen, S.~Parikh, I.~Saberi, B.~Stoica, and
  B.~Trundy, ``{Signs of the time: Melonic theories over diverse number
  systems},''
\href{http://arxiv.org/abs/1707.01087}{{\ttfamily arXiv:1707.01087 [hep-th]}}.

\bibitem{Bulycheva:2017ilt}
K.~Bulycheva, I.~R. Klebanov, A.~Milekhin, and G.~Tarnopolsky, ``{Spectra of
  Operators in Large $N$ Tensor Models},''
  \href{http://dx.doi.org/10.1103/PhysRevD.97.026016}{{\em Phys. Rev.}
  {\bfseries D97} no.~2, (2018) 026016},
\href{http://arxiv.org/abs/1707.09347}{{\ttfamily arXiv:1707.09347 [hep-th]}}.

\bibitem{Klebanov:2018nfp}
I.~R. Klebanov, A.~Milekhin, F.~Popov, and G.~Tarnopolsky, ``{On the Spectra of
  Eigenstates in Fermionic Tensor Quantum Mechanics},''
\href{http://arxiv.org/abs/1802.10263}{{\ttfamily arXiv:1802.10263 [hep-th]}}.

\bibitem{Cameron:1976}
P.~J. Cameron, \href{http://dx.doi.org/10.1017/CBO9780511662102}{{\em
  Parallelisms of Complete Designs}}.
\newblock London Mathematical Society Lecture Note Series. Cambridge University
  Press, 1976.

\bibitem{Polchinski:2016xgd}
J.~Polchinski and V.~Rosenhaus, ``{The Spectrum in the Sachdev-Ye-Kitaev
  Model},'' \href{http://dx.doi.org/10.1007/JHEP04(2016)001}{{\em JHEP}
  {\bfseries 04} (2016) 001},
\href{http://arxiv.org/abs/1601.06768}{{\ttfamily arXiv:1601.06768 [hep-th]}}.

\bibitem{Maldacena:2016hyu}
J.~Maldacena and D.~Stanford, ``{Remarks on the Sachdev-Ye-Kitaev model},''
  \href{http://dx.doi.org/10.1103/PhysRevD.94.106002}{{\em Phys. Rev.}
  {\bfseries D94} no.~10, (2016) 106002},
\href{http://arxiv.org/abs/1604.07818}{{\ttfamily arXiv:1604.07818 [hep-th]}}.

\bibitem{Ruelle:1989dg}
P.~Ruelle, E.~Thiran, D.~Verstegen, and J.~Weyers, ``{Adelic String and
  Superstring Amplitudes},''
\href{http://dx.doi.org/10.1142/S0217732389001970}{{\em Mod. Phys. Lett.}
  {\bfseries A4} (1989) 1745}.

\bibitem{Das:1989fq}
S.~R. Das, A.~Dhar, A.~M. Sengupta, and S.~R. Wadia, ``{New Critical Behavior
  in $d=0$ Large $N$ Matrix Models},''
\href{http://dx.doi.org/10.1142/S0217732390001165}{{\em Mod. Phys. Lett.}
  {\bfseries A5} (1990) 1041--1056}.

\bibitem{Sugino:1994zr}
F.~Sugino and O.~Tsuchiya, ``{Critical behavior in c = 1 matrix model with
  branching interactions},''
  \href{http://dx.doi.org/10.1142/S0217732394002975}{{\em Mod. Phys. Lett.}
  {\bfseries A9} (1994) 3149--3162},
\href{http://arxiv.org/abs/hep-th/9403089}{{\ttfamily arXiv:hep-th/9403089
  [hep-th]}}.

\bibitem{Gubser:1994yb}
S.~S. Gubser and I.~R. Klebanov, ``{A Modified c = 1 matrix model with new
  critical behavior},''
  \href{http://dx.doi.org/10.1016/0370-2693(94)91294-7}{{\em Phys. Lett.}
  {\bfseries B340} (1994) 35--42},
\href{http://arxiv.org/abs/hep-th/9407014}{{\ttfamily arXiv:hep-th/9407014
  [hep-th]}}.

\bibitem{Klebanov:1994kv}
I.~R. Klebanov and A.~Hashimoto, ``{Nonperturbative solution of matrix models
  modified by trace squared terms},''
  \href{http://dx.doi.org/10.1016/0550-3213(94)00518-J}{{\em Nucl. Phys.}
  {\bfseries B434} (1995) 264--282},
\href{http://arxiv.org/abs/hep-th/9409064}{{\ttfamily arXiv:hep-th/9409064
  [hep-th]}}.

\end{thebibliography}\endgroup

\end{document}